# PRECISION MEASUREMENTS OF THE HYPERFINE STRUCTURE IN THE $2^3P$ STATE OF $^3He$

Marc Smiciklas, B.S.

Problem in Lieu of Thesis Prepared for the Degree of

MASTER OF SCIENCE

UNIVERSITY OF NORTH TEXAS

May 2003

APPROVED:

David Shiner, Major Professor
Duncan Weathers, Committee Member
Sandra Quintanilla, Committee Member
Samuel E. Matteson, Chair of the Department of Physics
C. Neal Tate, Dean of the Robert B. Toulouse School of Graduate Studies

Smiciklas, Marc, Precision measurements of the hyperfine structure in the $2^3P$ state of $^3He$. Master of Science (Physics), May 2003, 44 pp., 3 tables, 15 illustrations, 12 references.

The unusually large hyperfine structure splittings in the $2^3P$ state of the $^3He$ isotope is measured using electro-optic techniques with high precision laser spectroscopy. Originally designed to probe the fine structure of the $^4He$ atom, this experimental setup along with special modifications I implemented to resolve certain $^3He$ related issues has made possible new high precision hyperfine structure measurements. Discussed are the details of the experimental setup and the modifications, including in depth information necessary to consider while performing these measurements. The results of these hyperfine structure measurements give an order of magnitude improvement in precision over the best previously reported values.

ACKNOWLEDGMENTS

This research would not have been possible without the financial support of the National Science Foundation, the National Institute of Standards and Technology, and the University of North Texas. The interest that has been shown to this project by a wide range of people has been greatly appreciated.

I thank the faculty and the staff at the University of North Texas for all the ways that they contributed to this project. Most importantly, I thank Dr. David Shiner for his excellent guidance and support that has helped me throughout this work.

TABLE OF CONTENTS

LIST OF TABLES

LIST OF ILLUSTRATIONS

# CHAPTER 1

## INTRODUCTION

The field of high precision measurements isn't necessarily the most glamorous endeavor, but certainly the importance of this work cannot be understated. These measurements are what test the theories that advance our understanding of the physical world to new plateaus. As anyone who performs these types of measurements can attest, the pursuit of higher precision is a formidable task indeed. Many challenges present themselves, and practically every detail must be carefully considered and understood. To simplify some of these factors, a convenient system is chosen to study. Helium is an ideal system for this. While the hydrogen atom (the simplest atom with its single electron) can be solved exactly by analytical means in the relativistic and non-relativistic limit, helium (two electrons) has no exact solution, even in the non-relativistic limit. Therefore, an infinite series of higher order approximations must be used to describe the helium atom [1]: thus experimental measurements are important to test the theory. My group has developed laser and electro-optic techniques for studying the atomic structure of helium. Previous work focused on the fine structure in the $2^3P$ state of the $^4He$ isotope [2]. However, these same techniques (with some innovative modifications due to special issues related to $^3He$) are readily applied to the unusually large hyperfine splittings in the $2^3P$ state of the $^3He$ isotope. This is very convenient since $^3He$ is a slightly more complicated system with its hyperfine interactions and also much less studied, both experimentally and theoretically, as compared to $^4He$ [3,4]. Thus, study of $^3He$ can serve as consistency check on both the theory and experiment developed for $^4He$. Also, as a potential application, a better understanding of the hyperfine interactions leads to precise predictions for nuclear size through the isotope shift [5]. This has been used to test nuclear theory for the helium nucleus [5,6] as well as for other nuclei using the corresponding heliumlike transitions in, for example, lithium [7], beryllium [8], and fluorine [9]. In this experiment, the hyperfine structure splittings in the $2^3P$ state of the $^3He$ atom are measured to new levels of precision.

The basic experimental setup used for the measurements of the $^3He$ hyperfine structure has already been established for the purpose of measuring the fine structure of $^4He$. The technique involves using an infrared laser to excite atoms that have been collimated into a beam and prepared in an initial state. This setup is discussed in detail in Chapter 3. For $^4He$, the fine structure in the $2^3P$ state is probed by exciting atoms up from the $2^3S$ metastable state. The fine structure of $^4He$ is illustrated in Fig. 1. It has been recognized that this same experimental setup can be used to probe the hyperfine structure in $^3He$ (shown in Fig. 2) as well. This is considered important for the $^4He$ measurements since a very useful validation of the experimental technique can be preformed with $^3He$. While $^3He$ hyperfine structure is not generally known very precisely, there is a hyperfine splitting in the 2S metastable state that is know to very high precision. If this is measured correctly with this experimental setup, it would lend a great confirmation of the reliability for this technique. However, satisfactory results have not previously been found since these measurements have been neither reliably nor economically feasible.

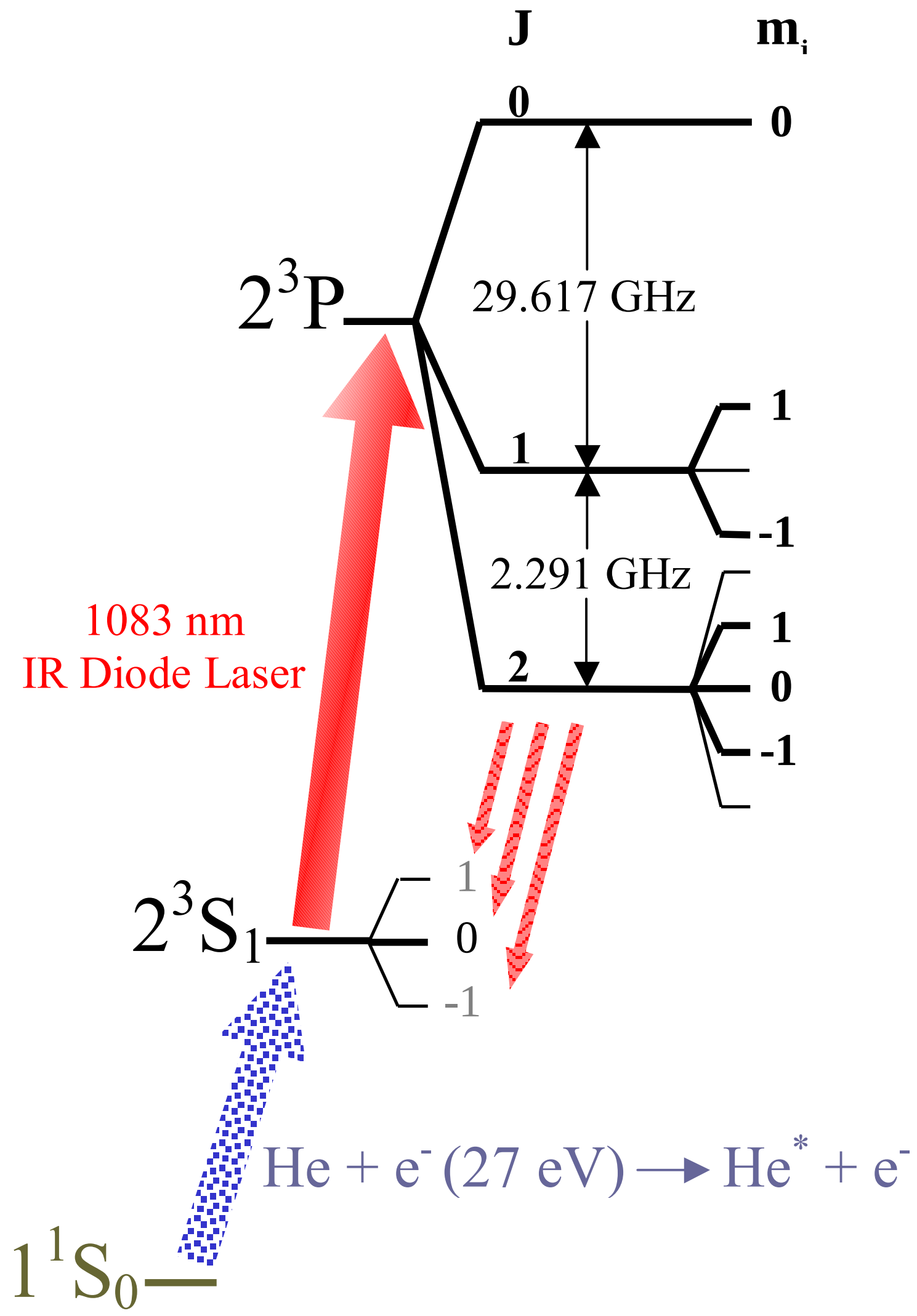

Fig. 1. $^4He$ Energy level diagram.

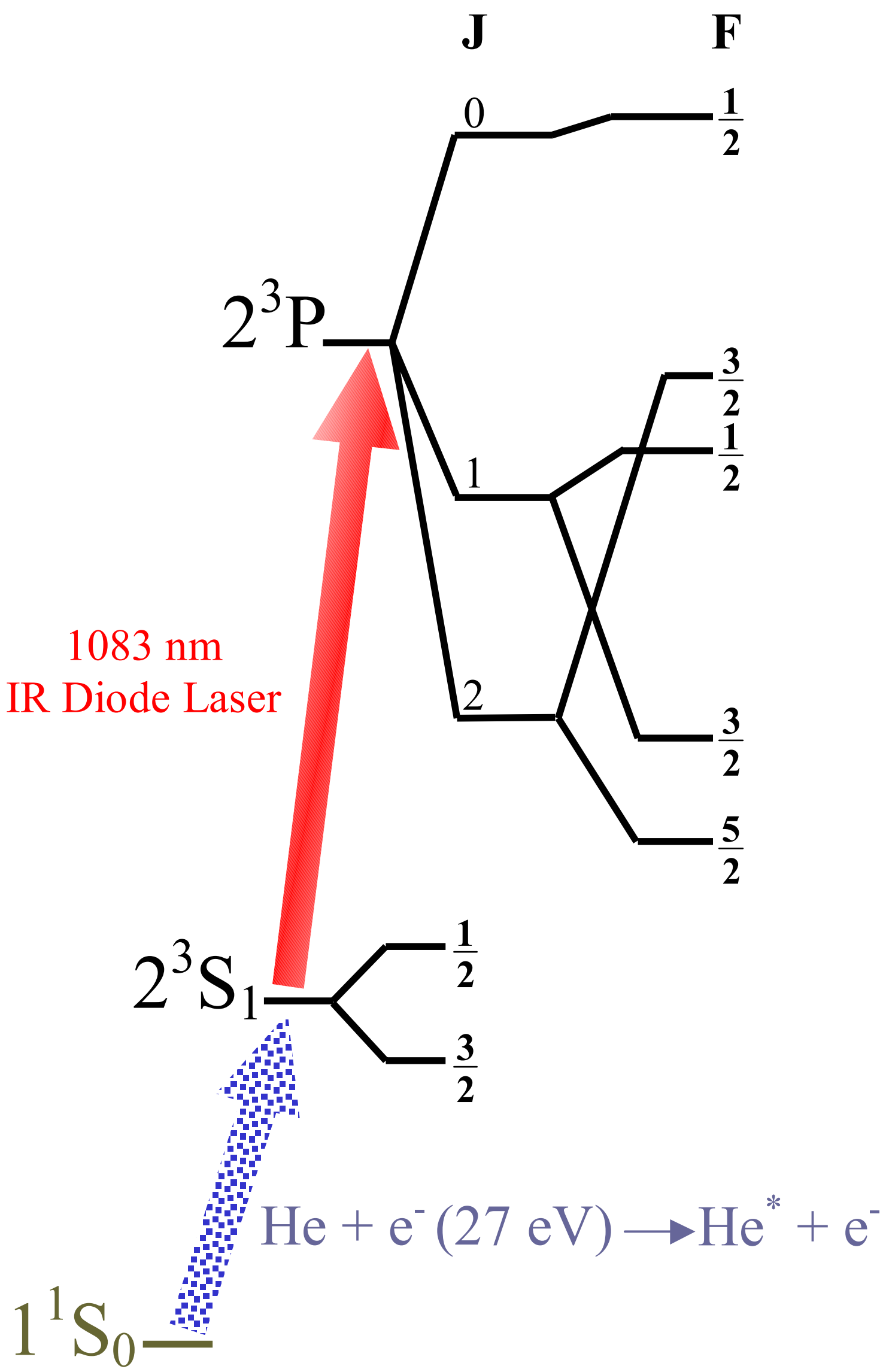

Fig. 2. $^3He$ energy level diagram.

During the process of preparing the initial state, unwanted atoms are deflected out of the atomic beam using a deflection magnet. The initial state in $^4$He experiences no deflection. However, $^3$He has two possible initial states, and they do experience small deflections. For the reliability of the experiment, this must be corrected. Also, $^3$He is not nearly as abundant as $^4$He. Therefore, it is quite expensive. The normal method for running an atomic beam experiment uses the helium gas far too rapidly. This is not an issue with $^4$He, but with $^3$He, this means that long runs, which are necessary to obtain the desired statistical accuracy, are not possible. Both of these problems have been address and solutions found, which are presented in this thesis along with the experimental setup. This has allowed high precision experiments, with precision comparable to $^4$He, to be conducted. With this setup, I report on the consistency check mentioned above and also measure the actual hyperfine splittings of the $2^3$P state of $^3$He.

# CHAPTER 2

# THEORETICAL FRAMEWORK

## Introduction

It is especially important when performing high precision measurements to understand in detail the system under investigation. Thus, a strong theoretical foundation must be established. Fortunately, this experiment has the benefit of a hundred years of quantum theory to help with that. By evaluating the quantum interactions that take place in the helium atom, we can not only understand its basic structure, but also reliably predict transition probabilities between the energy levels and, with the help of current helium theory, their approximate energy splittings. This foundation serves as a guide to our experiment when precisely measuring these same splittings. In fact, the results of this experiment ultimately serve as a test for these theoretical results.

In this chapter, the interactions that produce the various splittings in the first and second excited states (2S and 2P respectively) of the $^3$He isotope are introduced. By diagonalizing the resulting Hamiltonian operator, a new basis is obtained by which the transition probabilities between the relevant levels are then calculated. Consequently, this yields valuable information about the transitions such as excitation rates, decay rates into detectable states, repopulation rates into the same state, and ultimately the predicted signal size for our experiment. Finally, the approximate transition energies are calculated by utilizing current helium theory along with some empirical observations.

## Basic $^3$He Structure

For the purpose of observing the basic structure of the helium atom, we must find the fundamental quantum mechanical properties (i.e. the degrees of freedom which then lead to quantum numbers) of that system. Helium is an atom consisting of two electrons orbiting around a central nucleus. In the case of the $^3$He isotope, the nucleus contains two protons and a single neutron. Each of these particles has an intrinsic spin angular momentum quantum number associated with it. The two electrons in the system each have a spin angular momentum ($s_i$) equal to ½. However, the $^3$He nucleus can be treated as a single particle with total spin angular momentum ($I$) also equal to ½. Thus, the He-3 system consists of three spin-½ particles. The final quantum number to note is the orbital angular momentum ($l_i$) of each of the electrons. The value of this quantum number depends upon the orbital that an electron occupies. Each electron has an orbital angular momentum quantum number associated with it, and the total orbital angular momentum quantum number ($L$) is simply the sum of all the individual orbital angular momenta. For the purpose of this experiment, one of the electrons always occupies the ground state ($l_1 = 0$). Therefore, the total orbital angular momentum is simply equal to the orbital angular momentum of the outer electron ($L = l_2$). With these four quantum numbers and the basic rules for adding angular momentum, new quantum numbers can be found which appropriately describe the internal structures of the $^3$He atom. As will be seen, the radial quantum numbers can be limited, for the purposes of this experiment, to the 2S and 2P states. For illustration, refer back to the energy level diagram for these $^3$He states in Fig. 2 presented in Chapter 1.

The noninteracting part of the problem can be solved (i.e. solving the Schrödinger equation and obtaining the usual quantum numbers $n$, $l$, $m$, $m_s$) for each electron. The largest correction to this structure comes from the electron-electron Coulomb interaction, which gives rise to the singlet-triplet splitting. This splitting depends on the total spin angular momentum ($S$) of the system due to the spin of the electrons. By applying the standard method in quantum mechanics for addition of angular momentum [10], we see that the total spin angular momentum can take on a range of possible values given by

$$S = (s_1 - s_2), \ldots, (s_1 + s_2) = (\tfrac{1}{2} - \tfrac{1}{2}), \ldots, (\tfrac{1}{2} + \tfrac{1}{2}), \text{ in integer steps.}$$

So,

$$S = 0 \text{ and } S = 1 .$$

These are the only two possible total spin states for the helium atom. The $S = 0$ state is referred to as the singlet state, while the $S = 1$ state is termed the triplet state. The names are derived from the spin magnetic moment sublevels ($m_S$) which are enumerated by the standard prescription

$$m_S = -S, \ldots, +S \text{, in integer steps,}$$

thus yielding

$$\text{for } S = 0 \Rightarrow m_S = 0$$

and

$$\text{for } S = 1 \Rightarrow m_S = -1, 0, +1 .$$

Since in this experiment we use only the $2^3$S and $2^3$P states, we can focus our attention entirely to the case where $S = 1$, or in other words, the triplet configuration. Even at this level, the distinction of the spin magnetic moment sublevels is important to note. Their importance comes into play when preparing the atoms in their initial state, and again when detecting atoms that have undergone transitions. The key characteristic here is the deflection of the ±1 levels in a magnetic field gradient, and that the 0 level experiences no deflection.

Now that the total spin has been established, we can introduce the next structure into the system which involves finding the total angular momentum of the electrons ($J$), which is simply the addition of the total spin angular momentum to the total orbital angular momentum. This is the interaction that is responsible for the fine structure splitting. For the $2^3$S state, the orbital angular momentum is equal to zero ($L$ = 0), so here there is no fine structure splitting. However, for the $2^3$P state we have $L$ = 1. Thus, we can determine the fine structure levels in this state as follows

$$J = (S - L), \ldots, (S + L) = (1 - 1), \ldots, (1 + 1).$$

So,

$$J = 0,\ 1, \text{ and } 2.$$

Finally, we introduce the last interaction into the system, which is responsible for the hyperfine structure splitting. This involves adding the nuclear spin to the total angular momentum of the electrons to find the total angular momentum for the system ($F$). Recall that the nuclear spin is equal to ½, which consequently means that all levels are affected by this interaction. Now, following the same prescription as above, we have

$$F = (J - I), \ldots, (J + I) = (J - \tfrac{1}{2}), (J + \tfrac{1}{2}).$$

For the $2^3$S state,

$$J = 1,\ F = \tfrac{1}{2}, \tfrac{3}{2}.$$

For the $2^3$P state,

$$J = 0,\ F = \tfrac{1}{2};$$

$$J = 1,\ F = \tfrac{1}{2}, \tfrac{3}{2};$$

$$J = 2,\ F = \tfrac{3}{2}, \tfrac{5}{2}.$$

Therefore, we see that the $2^3$S state contains two hyperfine levels, and the $2^3$P state contains a total of five hyperfine levels. Each of these levels can again be split up into magnetic sublevels ($m_F$), or Zeeman levels, given by

$$m_F = -F, \ldots, +F.$$

The degeneracy in these levels is removed by placing the atom in a magnetic field. This produces a total of six sublevels in the $2^3S$ state, and a total of eighteen in the $2^3P$ state. Of the six sublevels in the $2^3S$ state, two are used as a possible initial state for exciting to the $2^3P$ hyperfine levels. These are the $F = 1/2$, $m_F = -1/2$ and $F = 3/2$, $m_F = 1/2$ magnetic sublevels, using the low field quantum numbers to label the states. These two levels correspond to the $m_S = 0$ spin magnetic moment whose energy has been split by the hyperfine interaction. Consequently, introducing the hyperfine interaction means that these two states are no longer relatively unaffected by magnetic fields. In fact, they now experience a small deflection by a magnetic field gradient. Due to this complication, special care has been taken in the experimental setup, which is discussed in detail in Chapter 3. For further illustration, a Breit-Rabi diagram for the $2^3S$ state is shown in Fig. 3.

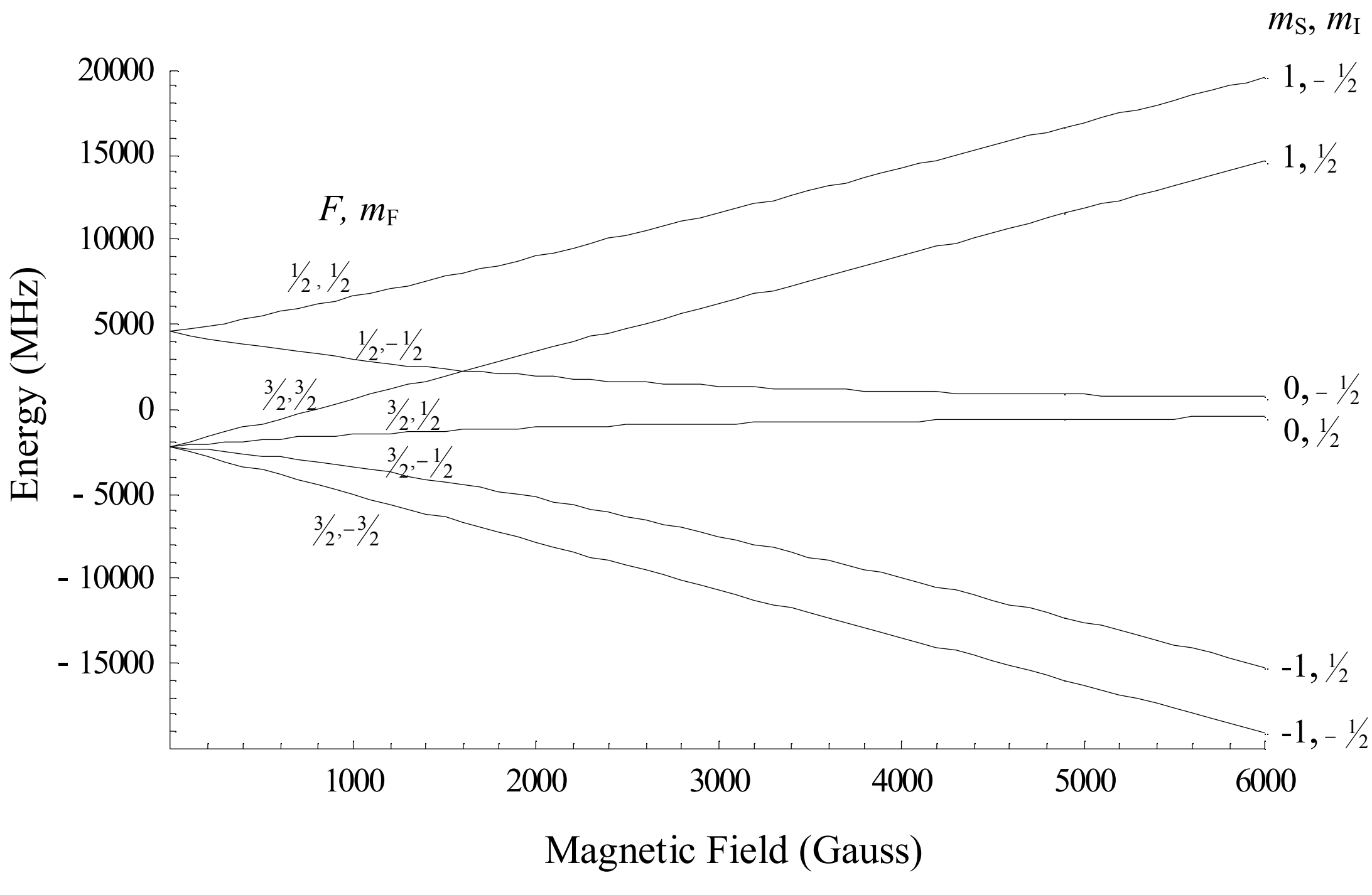

Fig. 3. Breit-Rabi diagram for the $2^3S$ state in $^3He$.

Hamiltonian Operators

In the previous section, the basic structure for the $2^3S$ and $2^3P$ states was established by adding together the various angular momentum quantum numbers. To extract more detailed information from the system, we must construct the Hamiltonian operators for the 2S and 2P states whose eigenvalues represent the energy for each level. Note that these are the complete Hamiltonian operators for each state, including both the triplet and the singlet levels. Here, the singlet levels cannot be ignored since several singlet-triplet mixing terms appear in the Hamiltonian that affects the triplet energies.

The Hamiltonian operator for the 2S state involves a spin-spin coupling between the electrons and a coupling between the electron spins and the spin of the nucleus. There is no orbital angular momentum interactions since $L = 0$ for S states. The Hamiltonian for the 2S state can thus be written as [4,6]

$$H_{2S} = \Delta S(\mathbf{s}_1 \cdot \mathbf{s}_2) + c(\mathbf{S} \cdot \mathbf{I}) + c'(\mathbf{K} \cdot \mathbf{I})$$

where

$$\mathbf{S} = \mathbf{s}_1 + \mathbf{s}_2 \text{ and } \mathbf{K} = \mathbf{s}_1 - \mathbf{s}_2 .$$

The operators $\mathbf{s}_1$ and $\mathbf{s}_2$ represent the spin operators for the two electrons. The operator $\mathbf{S}$ is the total electron spin operator, and the operator $\mathbf{K}$ is a singlet-triplet mixing operator. Spin-spin coupling is introduced by the $\mathbf{s}_1 \cdot \mathbf{s}_2$ term, which is responsible for the singlet-triplet splitting. The coefficient $\Delta S$ is adjusted to match the observed singlet-triplet splitting and can also be calculated theoretically to the required precision. The other two terms, $\mathbf{S} \cdot \mathbf{I}$ and $\mathbf{K} \cdot \mathbf{I}$, introduce the hyperfine splitting. These two terms are referred to as contact terms since they cause no energy shift for wave functions that are zero at the nucleus. The coefficients $c$ and $c'$ can be theoretically calculated given the electron wave functions.

The Hamiltonian operator for the 2P state involves all the same interactions as the 2S state; however, since $L = 1$ for P states, the Hamiltonian also includes orbital angular momentum coupling terms. The phenomenological form for the 2P Hamiltonian [4] is given by

$$H_{2S} = \Delta P(\mathbf{s}_1 \cdot \mathbf{s}_2) + E(\mathbf{S}\cdot\mathbf{L}) + c(\mathbf{S}\cdot\mathbf{I}) + c'(\mathbf{K}\cdot\mathbf{I}) + d(\mathbf{I}\cdot\mathbf{L}) + 2\sqrt{10}e\left(\mathbf{I}\cdot\left\{\mathbf{S}\mathbf{C}^{(2)}\right\}^{(1)}\right),$$

where

$$\mathbf{C}^{(2)} = \left(\frac{4\pi}{5}\right)^{1/2}\left[Y^{(2)}(\theta_1,\phi_1) + Y^{(2)}(\theta_2,\phi_2)\right]$$

is a tensor, and the curly brackets indicate the contraction of **S** with $\mathbf{C}^{(2)}$ to form a vector.

## Transition Probabilities

Now that the Hamiltonians for the 2S and 2P states have been established, some useful information concerning transitions between these states can be calculated. This is accomplished by determining the matrix elements of the laser perturbation $\sim\langle f|\mathbf{p}\cdot\mathbf{A}|i\rangle$, where $f$ and $i$ represent the final and initial states. In the electric dipole approximation this becomes $\sim\langle f|\mathbf{E}\cdot\mathbf{r}|i\rangle$. To determine relative transition probabilities, we can ignore overall factors and the radial part of the matrix element, which is identical for all transitions between the 2S and 2P manifold. Writing $\mathbf{E}\cdot\mathbf{r}$ in terms of spherical (as opposed to Cartesian) components, we obtain

$$T_q = \int_0^{2\pi}\int_0^{\pi} Y_1^{m_f}(\theta,\phi)\sqrt{4\pi}Y_1^{-q}(\theta,\phi)Y_0^0(\theta,\phi)\sin\theta d\theta d\phi .$$

This yields the usual selection rules for S to P transitions: $\Delta m = 0$ for the electric field polarized along the quantization axis and $\Delta m = \pm 1$ for right and left circularly polarized light. Of course, in our extended basis $|\psi(r_1,r_2), m_{S1}, m_{S2}, m_I\rangle$, we have the additional selection rules $\Delta m_{S1} = \Delta m_{S2} = \Delta m_I = 0$. This then defines the transition matrices $T_q$, which in the presence of the various perturbations above can be transformed to yield the relative transition probabilities between the new eigenstates of the system.

## Energy Eigenvalues

Determining the energy eigenvalues for the 2S and 2P states not only gives us the necessary information to calculate the transition energies between the levels, but also serves as the means by which the results of this experiment can be compared to theory. These relative energies are calculated by evaluating the Hamiltonians with the most advanced theoretically determined interaction constants. The best values to date have been calculated by Drake [6]. Differences between the evaluated Hamiltonian energies can be taken between the 2S and 2P levels to determine the theoretically predicted transition energies for the experiment. Also, differences within the 2P hyperfine levels can be taken to determine the theoretical hyperfine splittings. These values can then be compared to the results of this experiment to test the theoretical and experimental agreement.

## Magnetic Field Dependencies

The interactions in this experiment take place in the presence of an applied magnetic field. This magnetic field serves the purpose of maintaining the polarization states of each of the atoms. It also removes the degeneracy in the magnetic sublevels. Consequently, there is a relatively small energy shift to each of the levels. The experiment is conducted at a large enough magnetic field to minimize overlapping transitions. To illustrate this, Fig. 4 shows the transition energies of the magnetic sublevels for each of the hyperfine levels at a typical magnetic field value. Furthermore, it is crucial when performing the experiment and analyzing the data to understand precisely how each of these levels behaves in a magnetic field. Understanding this allows us to not only predict where the transitions are in a magnetic field, but also allows us to precisely measure the field. In the end, it is also necessary to remove the effects of the magnetic field by correcting the data to zero magnetic field. By doing this, we remove the magnetic sublevels and look at just the hyperfine levels that we are trying to measure.

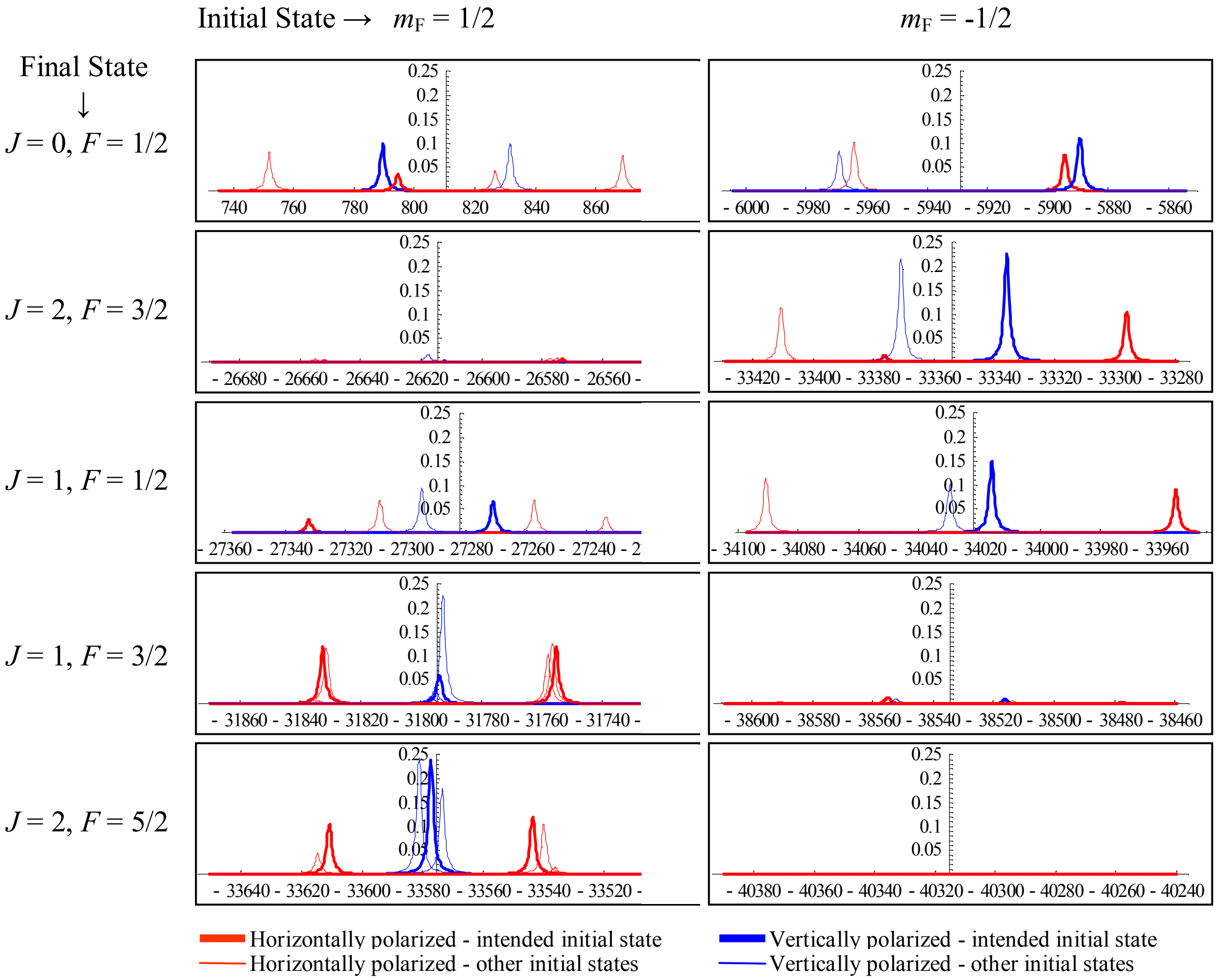

Fig. 4. Transition energy and signal size plots for relevant 2S to 2P transitions.

Calculating the magnetic field dependencies in the $^3$He atom requires modifying the Hamiltonians above to include magnetic field effects. This simply involves adding the terms along the quantization axis for the magnetic dipole moment [10]. Thus, the Hamiltonians can be written as follows,

$$H_{2S(B)} = H_{2S} + \mu_B B\left(g_I \mathbf{I}_z + g_S \mathbf{S}_z\right) \text{ for the 2S state,}$$

and

$$H_{2P(B)} = H_{2P} + \mu_B B\left(g_I \mathbf{I}_z + g_L \mathbf{L}_z + g_S \mathbf{S}_z\right) \text{ for the 2P state.}$$

The energy eigenvalues for these new Hamiltonians can now be found for any given magnetic field value. For convenience, the eigenvalues for a range of magnetic field values have been calculated, and a polynomial fit for each of the levels has been found. It is important to mention that these fits are used during the data analysis to correct the transitions to zero magnetic field. Therefore, the fits are taken to the $6^{th}$ order with residuals on the order of 1 Hz or better, which is 1000 times less than the targeted uncertainties. The reliability of these fits and the theory used to find them can be experimentally tested by verifying the consistency of the results at various magnetic fields.

# CHAPTER 3

# EXPERIMENTAL SETUP

## Introduction

A variety of techniques have been employed in this experimental setup. Some techniques are well establish and widely used in the scientific community, while others are more novel approaches that set this experiment apart from those groups doing similar measurements. Each of these techniques must be carefully considered and tested to insure their reliability and that, in fact, they have no adverse effects on the final results of the experiment. In this chapter, the essential techniques used are discussed for all major aspects of the experiment, along with specific noteworthy details that must be considered when doing the measurements.

## The Apparatus

### Overview

The apparatus, while seemingly simple in its basic setup, embodies most of the complexities that are involved in performing these high precision measurements. Many details must be pored over, and a significant amount of time invested into understanding the mechanisms at work here. A great deal of these mechanisms are understood, it is believed. However, there are some that are elusive. Some of these important details are discussed, along with the setup of each of the components in the sections of the apparatus. These sections comprise the preparation of the atoms into the initial state, exciting the atomic transitions, and detecting the atoms that have undergone the transition.

The apparatus is simply a large vacuum chamber that contains the mechanisms by which an atomic beam is created and prepared into the initial 2S state, then excited into a 2P state, and finally detected. Helium gas is leaked into what is referred to as the source region of the apparatus. After this, the atomic beam is formed by the first of two collimating slits and is sent through the interaction region and finally the detection region in which the beam is defined by the second collimating slit and detected. A partitioning wall placed immediately after the source region subdivides the apparatus into two chambers: the source chamber and the detection chamber. The atomic beam crosses this partition through the first collimating slit. Great care was taken to minimize the background gas in the detection chamber. By decreasing the background gas, the signal to noise can be increased during the detection process. Typically, the vacuum in the detection chamber is on the scale of $10^{-7}$ torr. See Fig. 5 for an illustration of the apparatus.

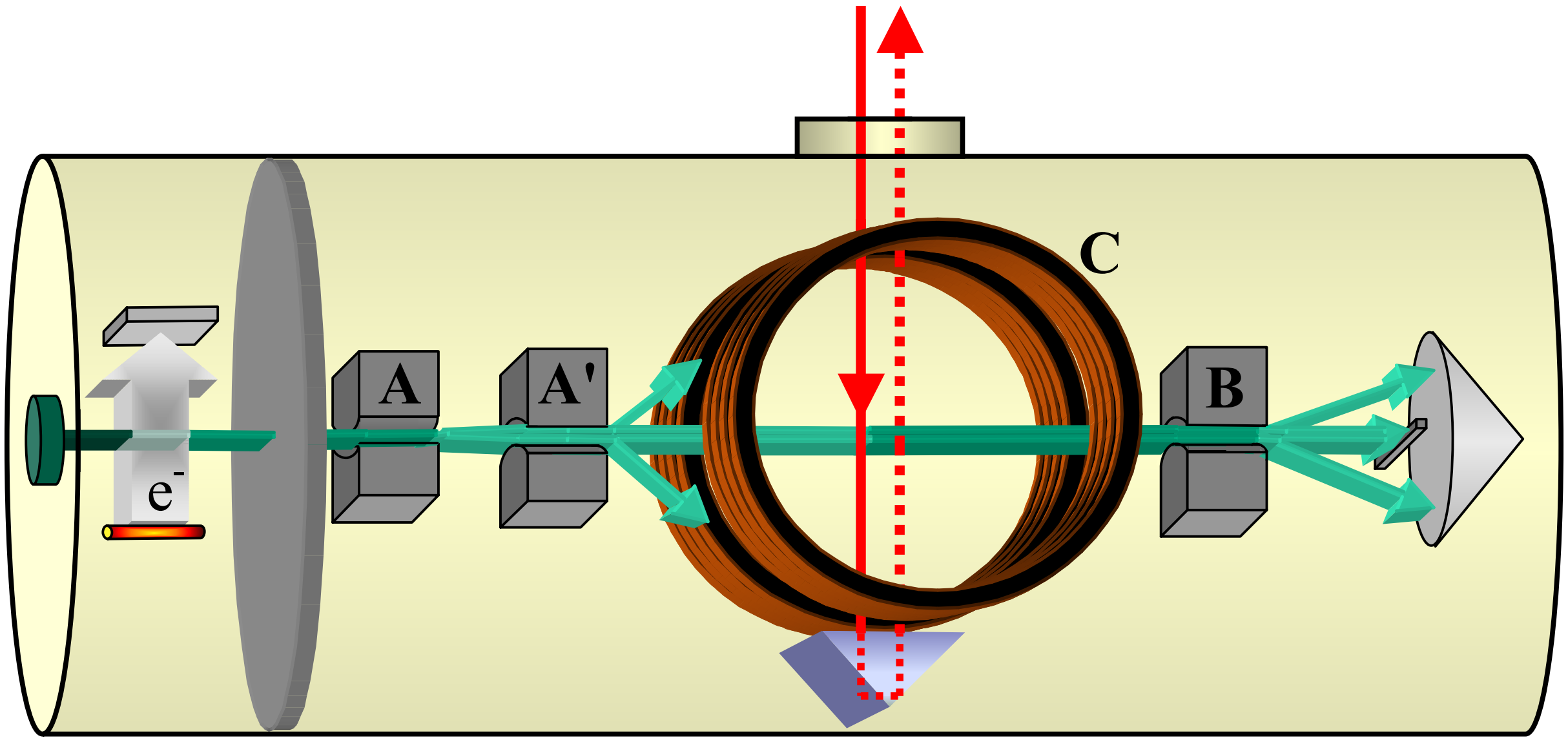

Fig. 5. Diagram of the experimental apparatus.

As has been stated in Chapter 1, there are economical constraints when working with $^3He$. As a way to conserve the gas and allow for long experimental runs, a recirculation of the gas was implemented. Initially, helium is injected into the source region above a hot tungsten filament that is used in preparing the initial state. After passing over the filament, the gas is formed into an atomic beam by the slit in the partition wall. Very little helium actually goes through the partition slit to form the atomic beam. Most of the helium disperses in the source region and is pumped away. Instead of just pumping it away, this new setup pumps the helium through a tube and reinjects it into the source region. However, the tungsten filament operates at very high temperatures to produce an emission of electrons. The helium going through the recirculation poisons this emission unless steps are taken to trap out contaminating gasses. These gasses could originate from an outside leak into the apparatus, or possibly from outgassing in the vacuum chamber. To trap the contaminants, the helium is sent through a copper tube which contains in a small portion of the tube a molecular sieve material that is submerged in liquid nitrogen. The very cold temperature of the liquid nitrogen allows the molecular sieve to trap out nitrogen, oxygen, and other contaminants. Oxygen, in this case, is believed to be primarily responsible for the poisoning of the emission. Quite surprisingly, the emission with this new setup is actually improved. The recirculation cleans up the helium gas in the experiment and is a better way to run. Such small amounts of helium gas are now used in the experiment that a small lecture bottle of $^3$He can now last for months of continual running, which before would have been depleted in less than a week.

Preparing the Initial States

To prepare for the excitation process, the helium atoms must first be prepared in the $m_F = 1/2$ and $m_F = 3/2$ or zero-state (referring to $m_S = 0$ for these two levels) magnetic sublevels of the 2S metastable state. This procedure is carried out by first putting the atoms in the metastable state by means of bombardment by electrons boiled off of the tungsten filament. Next, the other four levels (±1 states) of the $2^3$S state are deflected out of the beam by introducing a magnetic field gradient produced by a Stern-Gerlach type deflecting magnet (**A** magnet in Fig. 5). While it is true that the atoms in the singlet state remain in the atomic beam, they have little to no effect on the experiment. However, matters are complicated by the fact that the zero-states for $^3$He also experience a small deflection. Without some mechanism to correct for this deflection, the two levels have a Doppler shift between them of approximately 150 kHz. While it is possible to align the excitation laser and perform measurements from just one of the levels, it is uncertain what potential problems could arise from the direction change. Also, an important consistency check for the reliability of our experimental method would be eliminated, which is to compare the very precisely known value for the hyperfine splitting between $F = \frac{1}{2}$ and $F = \frac{3}{2}$ to a value that can be measured with this experiment.

In order to cancel out the Doppler misalignment between these two levels, a second Stern-Gerlach deflecting magnet (**A'** magnet) is utilized. To see how this works, we can use the results of the previous chapter where the energies of the atoms in the metastable state as a function of magnetic field have been determined. Now, the force is given by

$$\vec{F} = \frac{\partial E}{\partial x} = \frac{\partial E}{\partial B}\frac{\partial B}{\partial x}$$

For the deflecting magnets in this setup, the force is proportional to the magnetic field gradient, and the gradient scales approximately with magnetic field strength (to the extent that it is not limited by the saturation of the iron). Notice that at very high magnetic fields, the force felt by the zero-state atoms approaches zero. Therefore, deflecting the atoms with a large magnetic field to remove the ±1 states produces a small deflection in the zero-state atoms. Thus, by carefully tuning the second magnet to a very small magnetic field, the zero-state atoms can be counter deflected to cancel out the other deflection. While this is most likely not a perfect cancellation, any residual Doppler misalignment should be canceled out by a retro-reflecting prism implemented in the interaction region. This is discussed further in the next section.

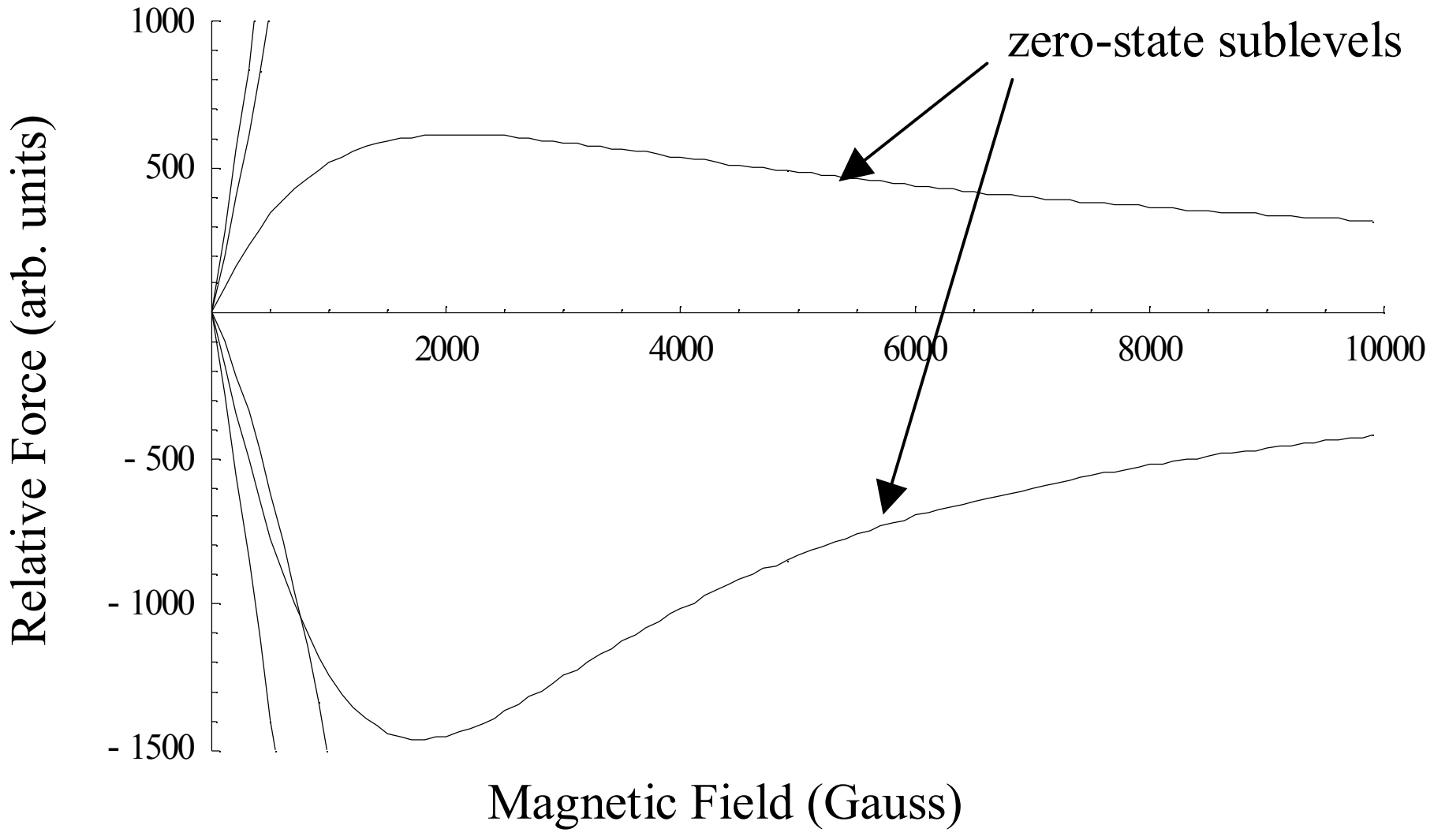

Fig. 6. Relative force on atoms in the $2^3S$ states in an inhomogeneous magnetic field.

Excitation to the 2P Hyperfine Levels

The excitation of the atoms to the 2P state takes place in the interaction region. In this region, the atomic beam traverses a uniform magnetic field that is created by a Helmholtz coil (**C** magnet in Fig. 5), where it interacts with an infrared laser beam via a side view port in the apparatus. The laser can be operated in essentially one of two modes by means of a retro-reflecting prism. The prism can be blocked for 1-way laser operation, or it can be unblocked for 2-way laser operation. Ideally, the 2-way laser cancels out all Doppler effects. However, due to losses in the retro-reflected beam caused by various factors, the typical Doppler cancellation is around 94%. Doppler effects play a very large role when considering the mechanisms involved in the interaction process. This and other effects, such as magnetic field and recoil effects, must be carefully considered when attempting to understand fully this stage of the experiment.

In order to consider carefully Doppler effects in this experiment, it is necessary to determine the velocity of the atomic beam. For this, a simple calculation can be carried out that gives the average velocity of the atoms by taking into consideration the equipartition of energy theorem:

$$\tfrac{1}{2}m\overline{v^2} = \tfrac{3}{2}k_B T \ ,$$

where, in this case, $m$ is the mass of the $^3$He atom, $k_B$ is Boltzmann's constant, and $T$ is temperature in degrees Kelvin. So, using $m = 5.01 \times 10^{-27}$ kg and $T = 298$ K, the average velocity of the atomic beam is calculated as follows:

$$\bar{v} = \sqrt{\frac{3k_B T}{m}} = \sqrt{\frac{3\times 1.38x10^{-23}\ J/K \times 298K}{5.01x10^{-27}\,kg}} = 1{,}570\,m/s$$

Now, if we consider that the atomic beam intersects the laser at precisely 90º, there is no Doppler shift in the frequency of the laser experienced by the atoms. However, if some small deviation in the alignment is introduced, a Doppler shift occurs. This Doppler shift is given by

$$\Delta f_{Doppler} = \frac{\bar{v}}{c} f \cos\theta = \frac{\bar{v}}{\lambda}\sin\theta_{dev} \ ,$$

where $\bar{v}$ is the average velocity calculated above, $\lambda$ is the wavelength of the laser (1083 nm), and $\theta_{dev}$ is the deviation angle from alignment. So, the Doppler shift for a small angle $\theta_{dev}$ is given by

$$\Delta f_{Doppler} = \frac{1570 m/s}{1.08x10^{-6}\,m}\sin\theta_{dev} \approx 1.45x10^{9}\,\mathrm{Hz}\cdot\theta_{dev}$$

While this is indeed a rather large Doppler effect, the laser in this experiment can easily be aligned to within 10 kHz. Also, the 2-way laser can be used to cancel out most of the Doppler misalignment and, in fact act as a consistency check when compared to the 1-way laser. It is important to note, however, that a Doppler misalignment should have no effect on the actual splittings between the hyperfine levels. Every transition has this same Doppler shift in energy, at least to the first order approximation. Therefore, they all shift the same amount, and the energy splittings are unaffected. There are subtle effects to consider such as direction biasing between the transitions. This occurs when one transition prefers a slightly different average atomic beam direction than another transition. An effect like this would mean that transitions can in fact have different Doppler shifts. This would affect the energy splittings. However, the 2-way laser should cancel out any Doppler shifts. Thus, there should be a discrepancy between the 1-way and 2-way results. Direction biasing between the transitions is discussed in more detail in the Detection section below.

Another important consequence of the Doppler Effect is the Doppler width of the atomic beam. Since the atomic beam has an average direction defined by the two collimating slits in the source and detection regions, there is a Doppler width in the atomic beam determined by the range of directions it consists of. The Doppler width can be calculated by the following formula,

$$\Delta f_{Doppler} = \frac{\bar{v}}{\lambda} \Delta\theta ,$$

where $\Delta\theta$ is the angular width determined by the separation of the collimating slits and their width. For the atomic beam in this experiment (0.1 mm collimation slit width separated by 0.40 m), the typical Doppler width is about 360 kHz.

Another effect to consider is the momentum transfer that takes place when an atom absorbs a photon. A photon carries an amount of momentum that is given by

$$p_{\gamma} = \hbar k = \frac{h}{\lambda}$$

In order to conserve momentum, the atom gets a small kick in the direction the photon is traveling. The atom also receives a kick when a photon is emitted, but that effect should average out since it is equally probable for the atom to emit the photon in any direction. However, the atoms that absorb a photon and undergo a transition will all receive a kick in the same direction.

Two effects must be realized when considering this. The first is that the excitation laser must not only supply the necessary energy to induce the transition, but must also supply the energy that goes into the momentum that the atom receives. This additional energy (represented in terms of frequency) is given by

$$\frac{p_\gamma^{\;2}}{2mh} \approx 56.4 \text{ kHz}$$

Fortunately, the amount of energy to supply to the momentum transfer is nearly the same for the 32 GHz range of transition frequencies in this experiment, and any additional energy is subtracted out of the splittings since it is essentially constant for all the levels. In fact, the maximum difference in momentum energy is ~35.4 kHz/$10^4$ and therefore negligible. The additional photon recoil energies need not be taken into consideration when determining the energy splittings. The second effect is that the average direction of the atomic beam is effectively changed, albeit a very small amount. The change in the average atomic beam direction due to photon recoil is $\sim p_\gamma / p_{He}$. This produces an additional Doppler shift given by

$$\Delta f_{recoil} = \frac{\Delta v_{recoil}}{\lambda} \approx 112.8 \text{ kHz}$$

One last detail to consider at this stage of the experiment is the magnetic field created by the Helmholtz coil. The coil generates a uniform magnetic field which causes the energy shifts in the magnetic sublevels during the excitation process. Field magnitudes up to 80 Gauss can be utilized in this setup. The magnetic field is designed to be uniform throughout the interaction region. It is possible that the atoms could sample an average magnetic field if gradients do exist. However, this could cause a problem if for some reason certain transitions sampled slightly different spatial regions than other transitions, and thereby slightly different magnetic fields. Many tests have been performed to see the effect that a magnetic field gradient would have on the results by introducing a strong magnet near the interaction region; however, no significant effects have been observed. Great care has also been taken to insure the stability of the magnetic field. This includes isolating the ground of the constant current source that supplies the coils to prevent noise pickup and especially cross talk between other devices in the experimental setup.

Detection

During the excitation of the atomic beam, atoms that have undergone a transition from one of the zero-states have some probability of decaying into the ±1 states. If no transitions have occurred, all the atoms remain in the zero-states. Therefore, we can detect the number of atoms in the ±1 states as a way of measuring the transition energies. Naturally, the closer the excitation laser is to the center transition energy, the more atoms there will be in the ±1 states and the larger the signal size will be. The detection of the ±1 states is accomplished by deflecting the atoms around a stopwire using a Stern-Gerlach deflection magnet (**B** magnet). Consequently, the atoms still in the zero-states hit the stopwire. Once around the stopwire, the atoms collide with an electron multiplier. The release of energy from an excited atom, which decays down to its ground state when it hits the electron multiplier, causes a cascade of electrons that result in a pulse being detected.

As was mentioned earlier, this method of detection does introduce the effect of direction biasing in the transitions. Although the effect is very small, transitions which decay primarily into one channel or the other (+1 states or -1 states) are biased to travel in a very slightly different path down the apparatus. This effect is caused by the ±1 states deflecting in opposite directions around the stopwire by **B** magnet. Atoms already heading in the direction that they will be deflected are more easily deflected around the stopwire, thus more easily detected. Transitions that decay equally into the ±1 states are not affected by this biasing. However, most transitions in $^3$He decay to one state more than the other. The width of the stopwire does affect the magnitude of this effect. Another stopwire-related effect is velocity biasing in the transitions. This effect is related to the centering of the stopwire. Since faster traveling atoms are deflected less, there is the potential for clipping of these atoms by the stopwire. Obviously, the centering of the stopwire will have opposite effects in terms of velocity biasing on transitions that decay primarily in one channel or the other.

## Optics

### Overview

Measuring various atomic transitions with a single laser demands the capacity to tune the laser frequency. For the $2^3$P state of the $^3$He atom, the laser must have a tunable range of approximately 32 GHz. This experiment uses a novel approach of implementing an electro-optic modulator (EOM) to create tunable side bands on the laser [11]. The EOM uses microwave frequencies to phase modulate a carrier frequency. Consequently, the frequencies of the side bands are tuned by inputting the necessary microwave frequency into the device. These side bands are then used to excite the transitions in the helium atom. However, if the frequencies and intensity of the side bands are to remain stable, the carrier frequency must remain stable. For this experiment, an infrared diode laser serves as the carrier frequency. Unfortunately, on its own, the diode laser neither outputs a very stable, well defined frequency nor does it exhibit a particularly steady intensity. Therefore, some mechanism for frequency and power stabilization must be implemented to control the laser. Otherwise, the side bands themselves will fluctuate along with the diode laser.

For the excitation laser in this experiment, a Distributed Bragg Reflector (DBR) diode laser is employed. The specific DBR laser implemented is particularly useful because of its broad range of frequency outputs around the wavelength 1083 nm, which is ideal for exciting transitions from the $2^3S$ to $2^3P$ states in helium. To select a more clearly defined frequency, the diode laser is mounted on the end of a one-meter resonant cavity. On the opposite end, a partially reflecting mirror fixed to a piezoelectric crystal is mounted for the purpose of precisely tuning the length of the cavity. By tuning the voltage on the piezoelectric crystal, and thereby altering the position of the end mirror, the resonant cavity forces the diode laser to lase in a mode whose wavelength closely coincides with the length of the cavity. Thus, a very narrow distribution of frequencies is transmitted. The transmitted laser is then split into two beams, one of which is used for frequency stabilization, and the other for the actual atomic excitations. The beam used to excite the atoms is first sent through an acousto-optic modulator (AOM) for the purpose of power stabilization, and then coupled into a fiber optic cable. The fiber optic cable makes it very convenient to move the laser beam wherever it is needed, due to the fact that no complex arrangement of alignment-sensitive mirrors is necessary to direct the beam. From there the beam is then passed through the EOM, and finally sent into the apparatus through a small porthole, after which it interacts with the atomic beam. Fig. 7 illustrates the optical components utilized in this experiment.

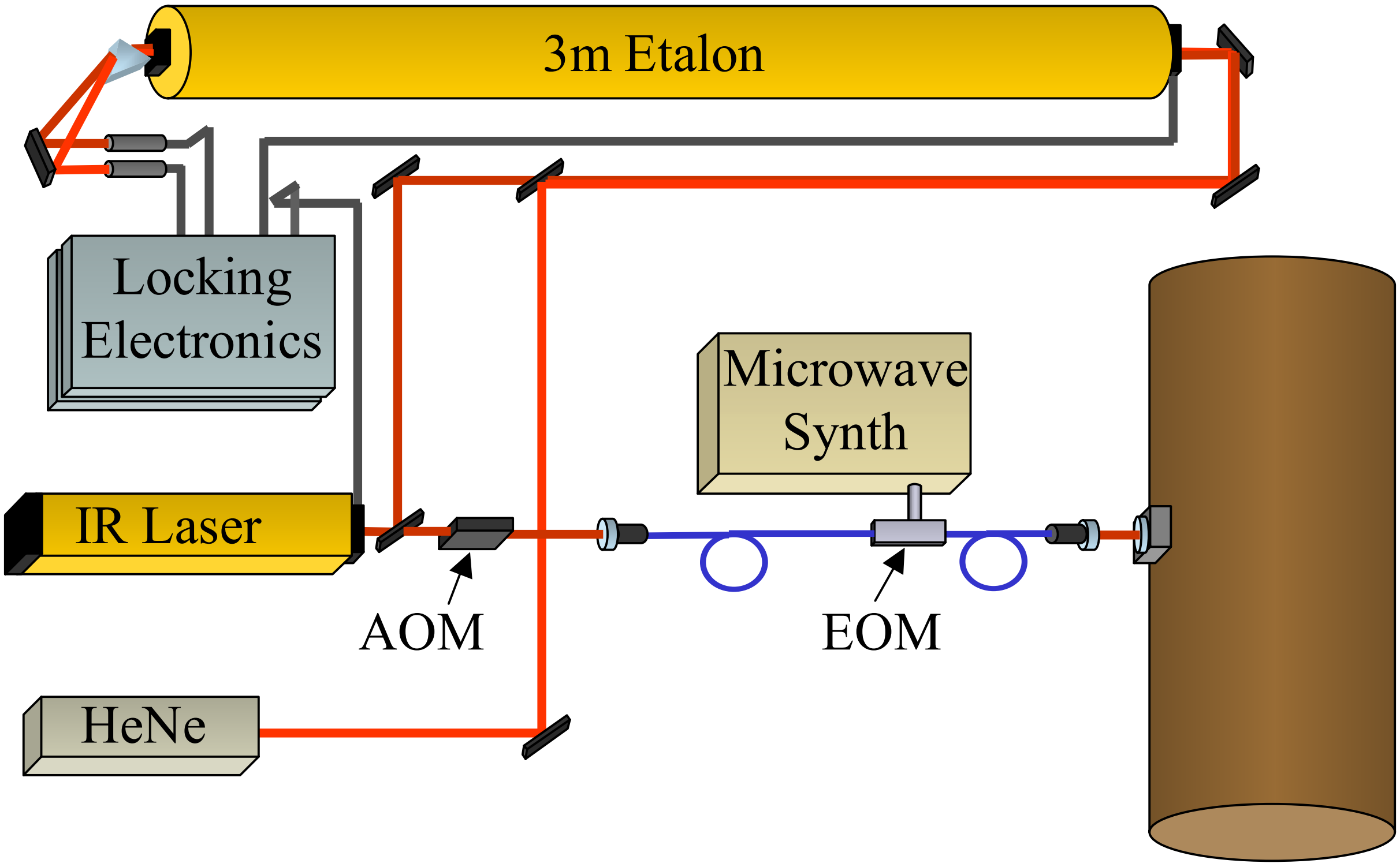

Fig. 7. Diagram of the optical layout.

Phase Modulated Laser Frequency Tuning

The phase modulation of a waveform induces side bands that are directly related to the frequency of the modulation. By taking a traveling wave with a time dependent phase, such as

$$\sin(\omega_c t + \phi(t))$$

where

$$\phi(t) = \beta \sin(\omega_m t),$$

and using of the appropriate trigonometric identities (assuming $\beta << 1$, which is sufficient for the purpose of this illustration), the traveling wave can be written as

$$\sin(\omega_c t) + \beta \sin[(\omega_c \pm \omega_m)t].$$

Here we have the original carrier frequency $\omega_c$ with side bands at intervals equal to the modulation frequency $\omega_m$ above and below the carrier. This is the same familiar result found for frequency modulation. For a more accurate representation of the frequencies that include higher order side bands than that presented above, an infinite series is necessary that incorporates the Bessel function and takes into account the modulation index (so we can no longer say $\beta << 1$).

The EOM in conjunction with a microwave synthesizer performs the modulation of the IR laser. It accomplishes this by modulating the index of refraction of a Lithium Niobate crystal. The modulation of the index of refraction effectively modulates the phase of the IR laser, which passes through a portion of the crystal that has been diffused with titanium to act as a waveguide. An optimal modulation index ($\beta$ = 1.84) is used to modulate the carrier frequency. At this setting, the intensities of the first order side bands become insensitive to small changes in microwave power output. However, the corresponding microwave power to produce this modulation index varies depending on the microwave frequencies. Fortunately, the necessary microwave power is relatively constant in time, which means that the required microwave powers can be mapped out according to frequency.

Laser Frequency Stabilization

The IR laser diode on its own does not posses the frequency stability necessary for high precision measurements. Thus, the need for frequency stabilization is critical. To stabilize the IR laser, its wavelength must be locked to a reference which does possess the desired stability. For this, an iodine-stabilized helium neon (HeNe) laser is utilized. A 3-meter long resonant cavity (3m etalon) is locked on to an integer number of wavelengths of the HeNe laser. Thus, the 3m etalon's length takes on the stability of the HeNe laser. The IR laser is then locked onto the 3m etalon, thereby transferring the stability of the HeNe laser onto the IR laser.

Power Stabilization

The laser power is stabilized using an acousto-optic modulator (AOM). The AOM diffracts the laser beam into $0^{th}$ order, $1^{st}$ order, $2^{nd}$ order, and so on diffracted beams. The $1^{st}$ order diffracted laser beam is then coupled into a fiber optical cable. From here it is sent to be phase modulated and then sent into the apparatus. Immediately before entering the apparatus, a small portion of the laser is split from the beam which is used to monitor the power and control the AOM. The AOM diffracts the laser beam by sending a high frequency acoustic traveling wave through a crystal. High and low density regions of altered index of refraction are created in the crystal by the acoustic wave's peaks and the troughs. This creates an effective diffraction grating for the laser. The intensity of the diffracted beam is dependent upon the power of the input acoustic waves. This is controlled by a closed loop negative feed back mechanism based on the monitored laser intensity. As a side note, it is true that the acoustic waves sent through the crystal are traveling waves. This in fact creates a Doppler shift in the laser when it diffracts off of the acoustic wave fronts. However, this detail is unimportant since in this experiment knowledge of the exact value of the carrier frequency is unnecessary.

## Data Collection

### Overview

To accurately determine the results of the experiment, all the necessary data must be collected. This not only includes the data to find the center of the transition energies, but also data must be collected to correct for power shifts and to determine an accurate value of the magnetic field in the interaction region. All these data must have sufficient statistical certainty to achieve the desired level of overall observational certainty. For this to be feasible, a computer-automated program is used to control the parameters of the experiment and collect all the data. The program can collect the data over long periods of time until the desired level of uncertainty is reached. At this point, the data can be analyzed by the computer to determine the results of the experiment.

### Frequency Stepping

The center of each transition is determined by analyzing data points taken on either side of the transition's energy distribution. Two points of data are taken on each side of the transition to determine the two slopes. These slopes along with overall signal size of the data points are then used to calculate the center transition energy. By doing this, the data can be collected at points on the transitions which contribute most to minimizing the statistical uncertainties. Other experimental groups doing similar measurements typically find a resonance curve which involves take data over the entire transition. The disadvantage to this is that too much time is spent on data points that contribute little to minimizing the statistical uncertainties and not enough time on the more important points. Of course there could be some concern with our method about the shape of the distribution and how symmetric it is. Systematic checks at various step sizes from the center of the distributions can be performed to verify the validity of our approach. Fig. 8 shows the Doppler and saturation broadened Lorentzian distribution of the resonance curve (solid line) for a helium transition (the dashed line is a simple Lorentzian). It also illustrates the method by which the data is collected using the pairs of data points.

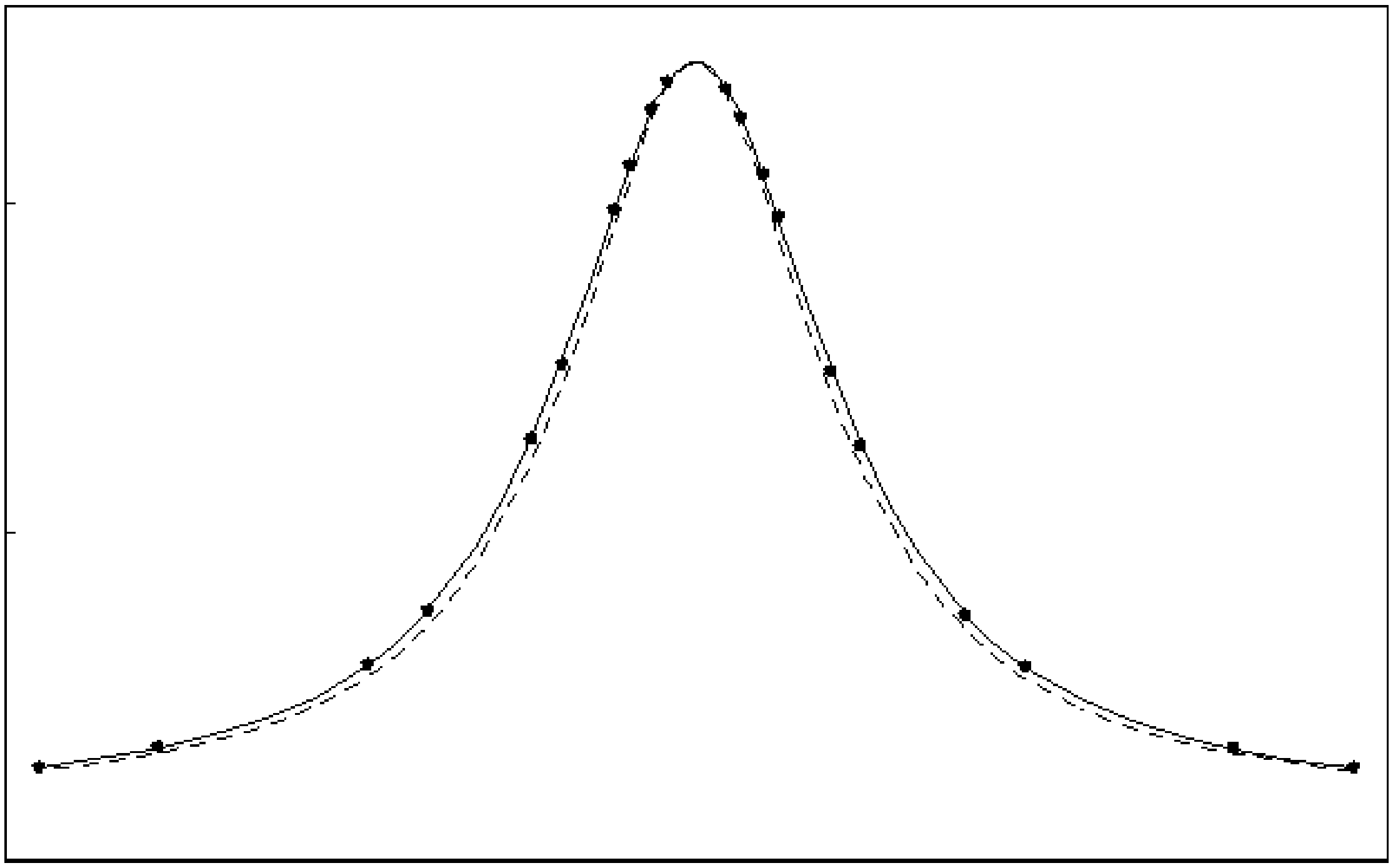

Fig. 8. Resonance curve for a helium transition fit to a Lorentzian.

Power Extrapolation

As the intensity of the excitation laser is increased, the energies necessary to drive the transitions shift. This shift in frequency is not necessarily the same for all transition. Therefore, the power shift affects not just the transitions but the splittings in the hyperfine levels as well. Because of this, the transition energies must be extrapolated to zero power to get the correct splittings. To extrapolate to zero power, the data collection program steps the intensity of the laser to take data at high and low power. The program controls the intensity of the laser via the AOM described above in the Power Stabilization section.

There are a number of possible reasons for the power shift, and in fact, the slope of the power shift for the one-way laser is very different than that of the two-way laser. Primarily, it is believed to be caused by atoms that undergo a transition more than once. At low power, the probability of an atom undergoing a transition more than once is very small, but as the power increases, it becomes more likely. The reason this causes a power shift is actually due to the momentum kick the atom gets from the first photon which changes its direction and the Doppler shift from the second photon. The two-way laser is more complicated and not entirely understood, but it is believed to be a combination of the momentum kick and the fact that the incident laser and the retro-reflected laser are overlapping each other, which allows absorption and stimulated emission from counter propagating photons.

Measuring the Magnetic Field

The magnetic field is measured by using two magnetic sublevels of the same hyperfine level. Since at zero magnetic field these levels are degenerate, the actual magnetic field in the interaction region can be measure by calculating the necessary field that would cause the splitting observed in the sublevels. $^3$He has two feasible hyperfine levels in the 2P state that can be used for this: $J = 1$, $F = \frac{3}{2}$ and $J = 2$, $F = \frac{5}{2}$. As consistency check, the two measured magnetic fields can be compared to each other; and even better, they can be compared to the magnetic field measured with $^4$He.

Statistical Accuracy

During the analysis of the data, the computer calculates the errors in the data and uses error propagation to determine the errors in the final results. There are two types of errors that it calculates: external errors and internal errors. The external errors only take into account the $\sqrt{N}$ noise based on the number of counts (N) collected. Thus, these errors are reduced by conducting longer experimental runs (and also with better signal to noise.) The internal errors actually show the internal consistency of the data itself (i.e. repeatability which can be found by looking at the fluctuations in the data during the run). Effects such as atomic beam instability, frequency jitter, and magnetic field jitter all have an effect on these errors. When the internal errors and the external errors agree (typically they agree to better than 10%), then repeatability is only affected by $\sqrt{N}$ counting fluctuations. In this scenario, it is possible to focus our attention entirely on systematic errors (as opposed to random errors) in the experimental setup or the way the data are being taken.

Long data runs are necessary for obtaining the desired statistical accuracy in the experiment. The implementation of the recirculation discussed in the setup of the apparatus is imperative for this reason. By running with recirculation, the experiment can be conducted over extended periods of time (typically overnight). This combined with the automated setup of the data collection is truly what enables the collection of the amount of data necessary to produce the new levels of precision for these $^3$He measurements.

# CHAPTER 4

# DATA ANALYSIS AND RESULTS

## Introduction

There are a variety of checks that can be performed to insure the reliability and accuracy in the results of the data. These checks can be categorized in one of two groups: consistency checks and systematic checks. Consistency checks compare results for which we know what the outcome should be. For instance, two possible ways of measuring the same interval should yield the same result. If this is not the case, then clearly there is a problem. Also, there is a very important "external" consistency check that involves comparing a very precisely known value for a hyperfine splitting in the 2S state to a value that we can measure (albeit by a rather unusual approach). As for systematic checks, these basically involve turning every knob available to see if it affects the results. If nothing seems to change the answers (at least in ways that can't be explained), then either the answer is right, or all the knobs haven't been found to turn. Finally, after all the checks are complete and the data have been thoroughly analyzed, the final results can be quoted.

## Consistency Checks

### 2S Hyperfine Splitting

The first consistency check to examine is the "external" consistency check mentioned above. This involves measuring the same $2^3P$ hyperfine level using both of the metastable zero-states, and then taking the difference to find the energy splitting between those metastable states. This is obviously not the best way to measure this 2S splitting, but for the purposes of this experiment, it can serve as a very valuable consistency check. If the results of this test are correct, it lends a significant amount of credibility to the experimental technique and the reliability of the other results. The measured result for the 2S hyperfine splitting can be compared to the very precisely known value of 6739.701177(16) MHz [12]. This is actually the most difficult measurement for the entire experiment. In fact, this measurement is one of the primary reasons for all the effort that was taken to correct the deflection problem when preparing the initial states. Consequently, it can only be reliably preformed using the 2-way laser to cancel out any residual Doppler misalignment between the zero-states. Data for various conditions are shown in Fig. 9 with $\pm\sqrt{N}$ counting errors displayed (0 kHz represents the well known value quoted above). In this plot and all subsequent plots the error bars indicate $\pm 1\sigma$ in the counting uncertainty. Therefore, they do not include any other contributions to the random uncertainty or any systematic errors.

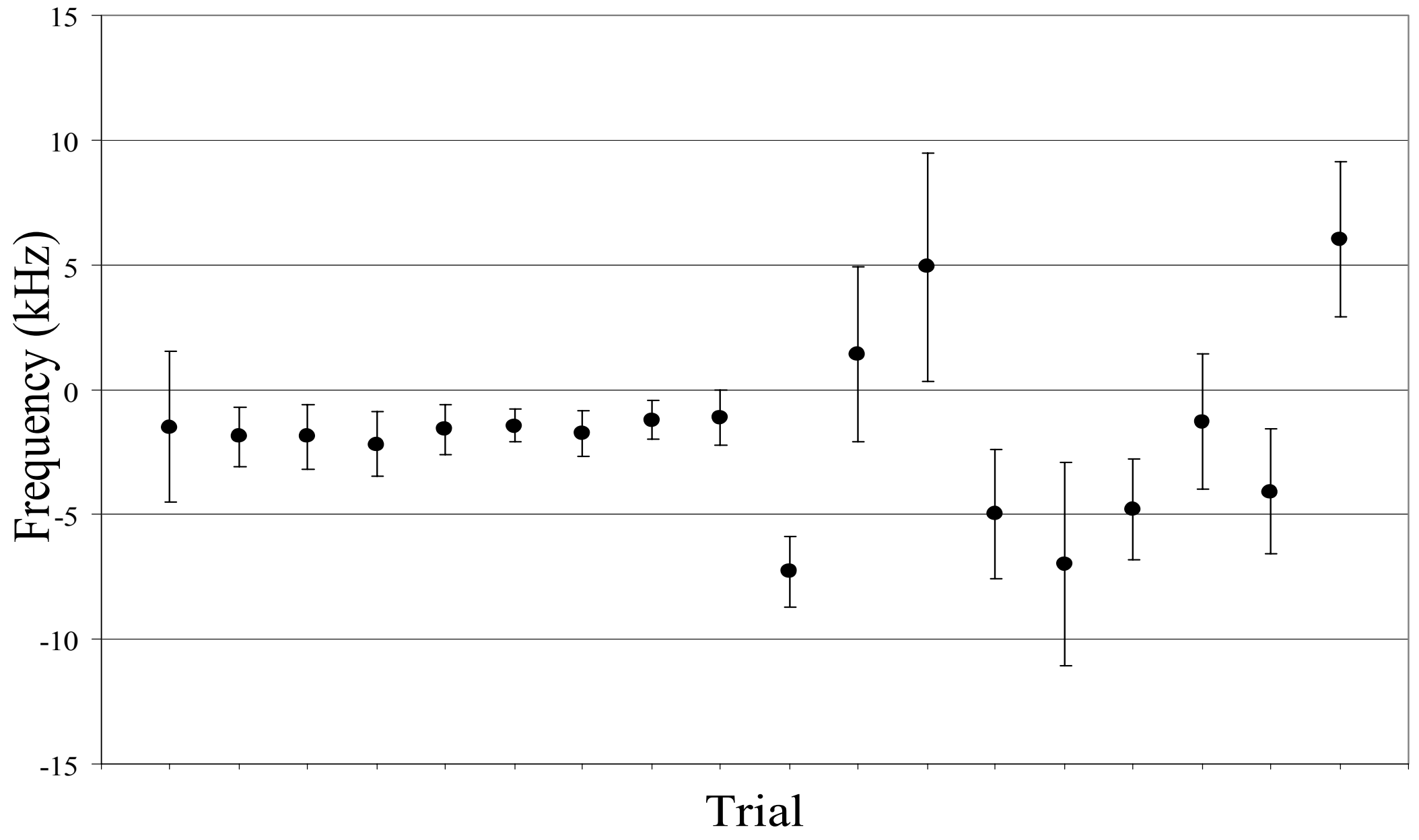

Fig. 9. Trial data for the 2S hyperfine splitting consistency check.

Notice the ~2 kHz offset and especially the spurious data towards the end in the figure. A variety of systematic checks were preformed, and most notably, the stopwire alignment seemed to cause inconsistency in the results. This is interesting since the stopwire alignment does not seem to have a significant effect on the $^4$He measurements performed by this group. It is my belief that the balance by which the atoms decay into the detectable states is what is behind this effect. $^4$He atoms decay either almost entirely to one of the detectable channels, or they are very nearly equally split between the channels. However, $^3$He does not exhibit this type of symmetry in the detectable states. Thus, any directional biasing that occurs caused by the stopwire could potentially have adverse effect on the results. This is a difficult problem to quantify since there is some uncertainty in what the "correct" stopwire alignment should in fact be. When looking at this problem, the stopwire is misaligned enough to create a calibrated increase in background counts detected. Clearly, this effect must be carefully considered when performing the systematic checks on the measured $2^3P$ hyperfine levels.

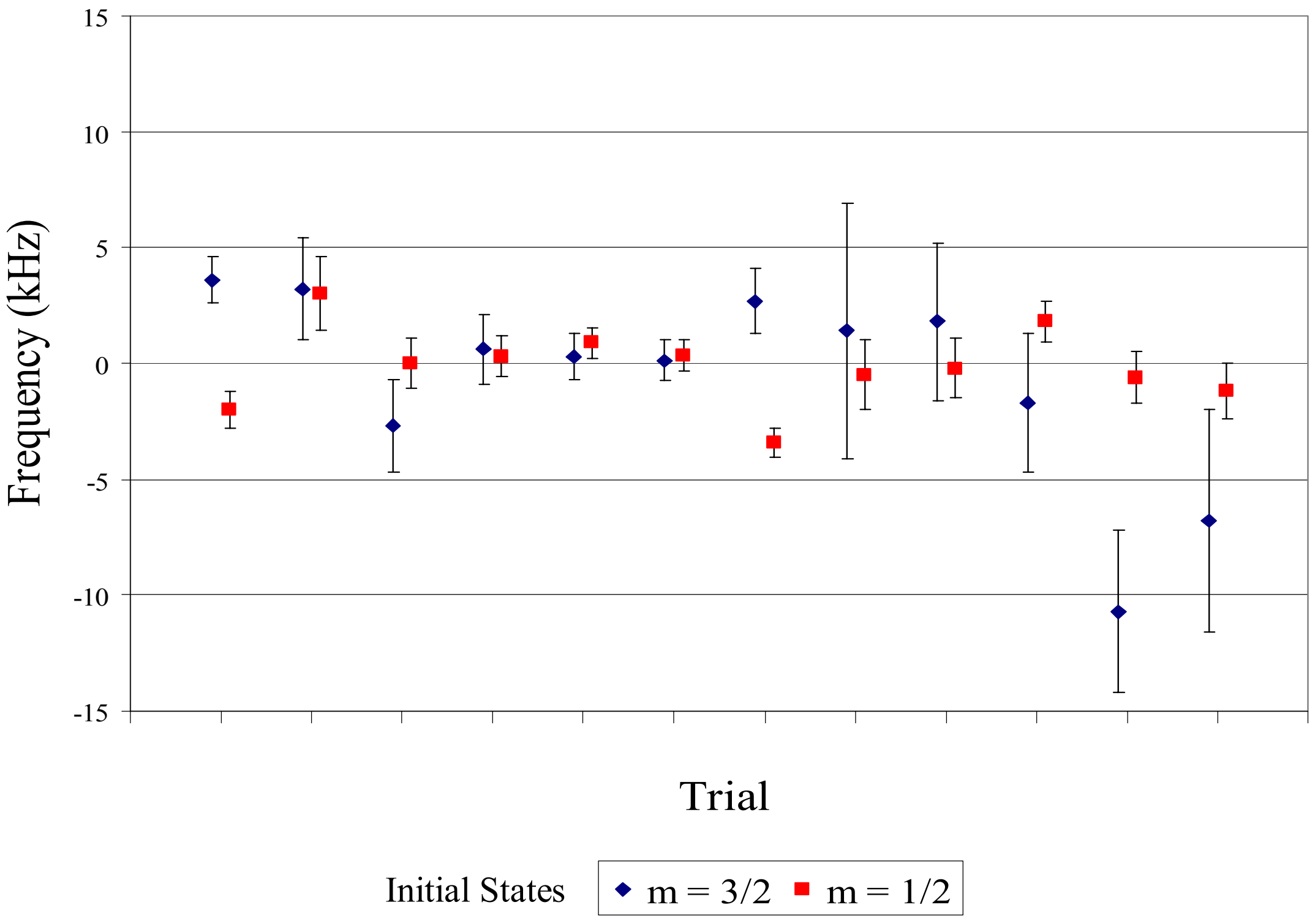

Fig. 10. Trial data for the J = 0 to 1 consistency check.

$J = 0$, $F = ½$ to $J = 1$, $F = ½$ Hyperfine Splitting

The splitting between $J = 0$, $F = ½$ and $J = 1$, $F = ½$ can be measured using either of the metastable zero-states. Therefore, the results of these two measurements can be compared for consistency. Results (without systematic checks) are given in Fig. 10 for both of the initial states. Paired data points were taken in the same trial run.

Magnetic Field Measurements

As described in the previous chapter, the magnetic field can be measured with two hyperfine levels in $^3$He: $J = 1$, $F = 3/2$ and $J = 2$, $F = 5/2$. Also, the experiment can be set up to run both $^3$He and $^4$He simultaneously. So, the magnetic field can be measured with $^4$He as well. These values can then be compared as a consistency check for the magnetic field. Results for this test are given in Fig. 11, where the $J = 1$ and $J = 2$ levels are plotted. The plotted values are relative to the measured $^4$He magnetic field in that trial run. Paired data points are from the same trial run.

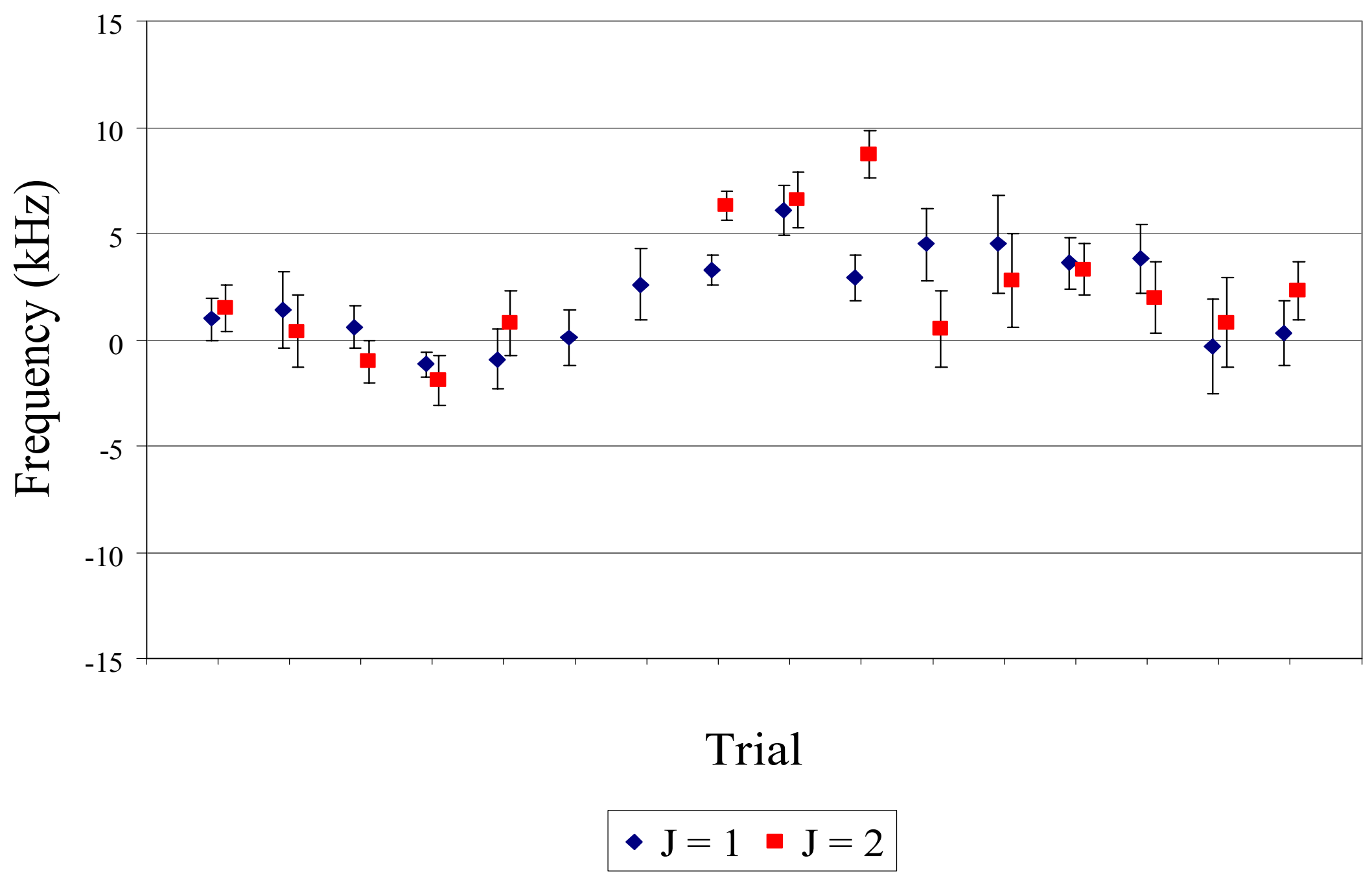

Fig. 11. Trial run data for the magnetic field consistency check.

## Systematic Checks

### Doppler Tests

By intentionally misaligning the excitation laser with respect to the atomic beam, Doppler shifts can be introduced into the transition energies. To the first order approximation, there should be no effect on the hyperfine splitting since all levels experience the same Doppler shift. However, other effects such as directional and velocity biasing could appear during these tests. The 2-way laser is especially important for these tests since it cancels out a significant portion of the Doppler shifts. Results using the 2-way laser at very large Doppler misalignments are shown above in Fig. 12. Clearly, even with large misalignments, the effects are minimal.

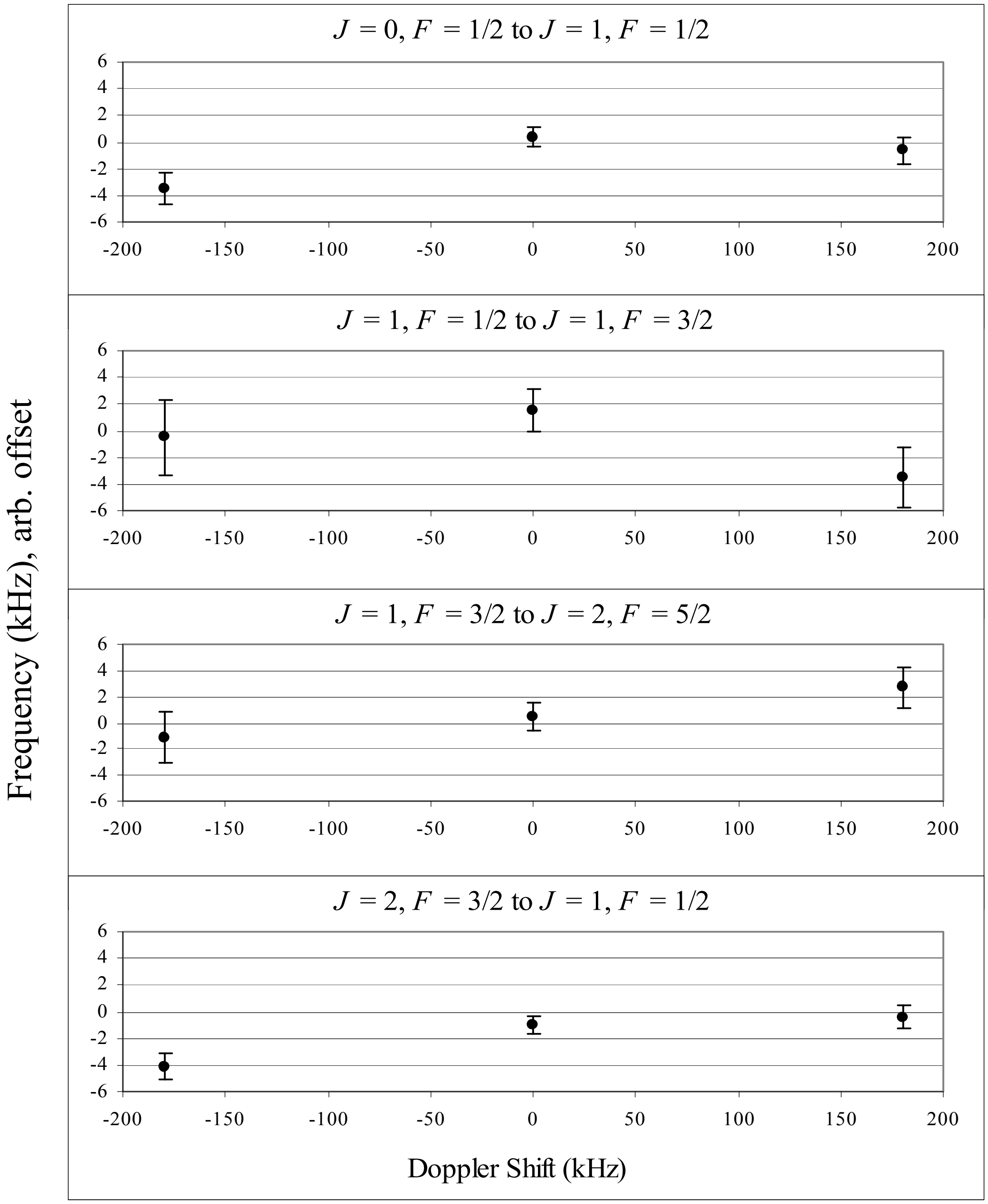

Fig. 12. Measured Doppler check data for the $2^3$P hyperfine splittings.

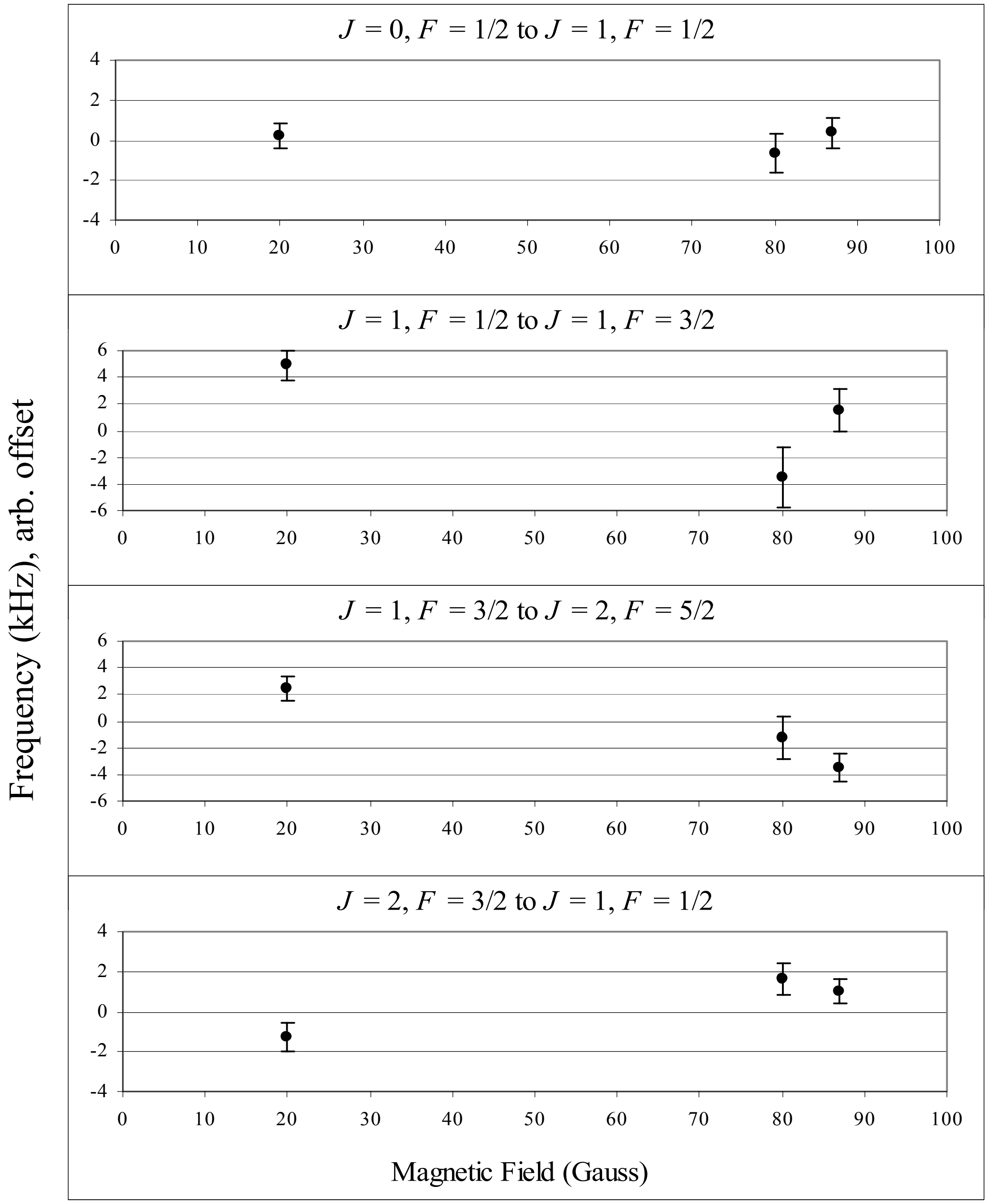

Fig. 13. Measured magnetic field data for the $2^3P$ hyperfine splittings.

Magnetic Field Tests

It is important to perform systematic checks related to the magnetic field. One reason is to test the validity of the magnetic field corrections. Also, spatial irregularities such as magnetic field gradients could potentially introduce errors. These spatial irregularities can presumably be altered by changing the magnitude of the magnetic field. Fig. 13 shows the results of these tests.

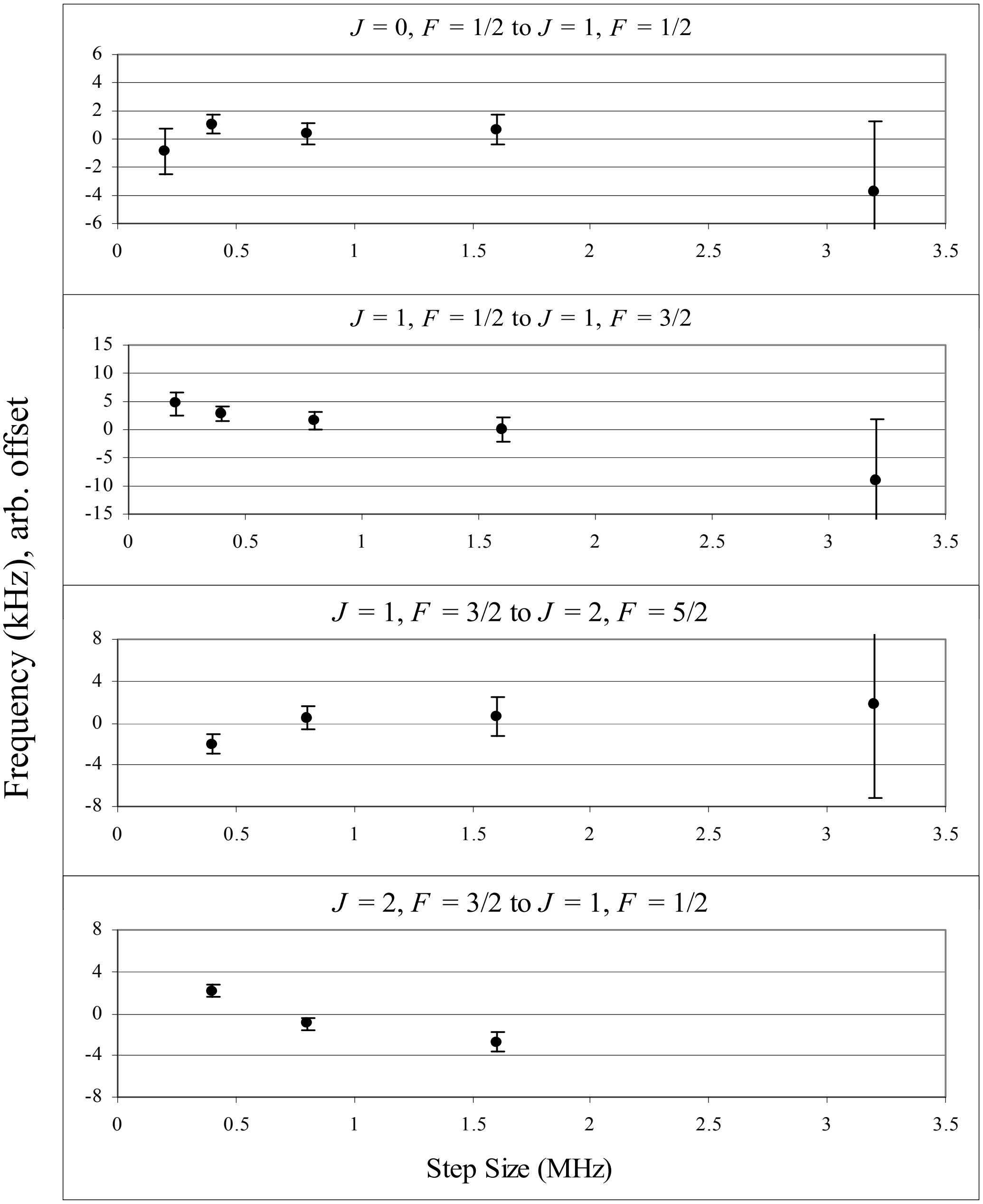

Fig. 14. Measured step size data for the $2^3$P hyperfine splittings.

Frequency Step Size

While I believe the method of data collection used in this experiment offers significant advantages over resonance curves, the possibility of asymmetries cannot be ignored. Multiple tests have been conducted at various frequency step sizes from the center of the transition distributions. If in fact asymmetries exist, these tests should reveal them, see Fig. 14.

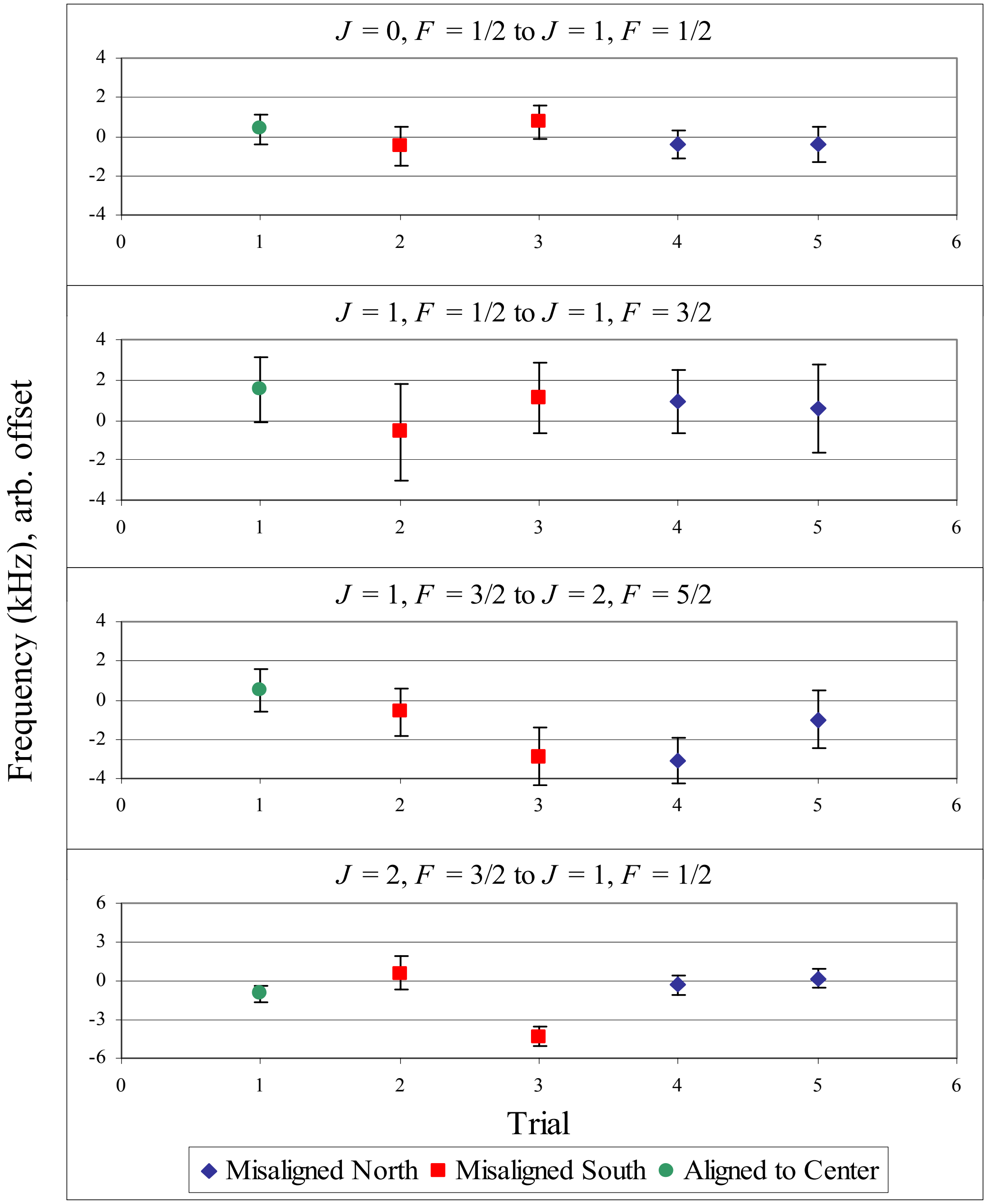

Fig. 15. Measured stopwire misalignment data.

Stopwire Alignment

Due to the effects of directional and velocity biasing that occur in the detection process, it is important to look at how stopwire alignment affects the measured separation energies for various hyperfine splittings. Several tests were performed with stopwire misalignments in both directions, referred to as north and south misalignments. The results are shown in Fig. 15.

## Experimental Results

The systematic checks shown above and the various other tests that have been performed give me confidence in the results of this experiment. Clearly the results for the Doppler tests and magnetic field test show that the data are not adversely affected to any significant degree by these effects. The poor results for the data taken at large step sizes from the center of the transitions is not, in my opinion, a cause for concern. At very large step sizes, the data are less statistically certain due to smaller signal sizes with respect to the noise. Also, the potential for overlapping transitions to affect the data becomes much more significant. As for the stopwire tests, the effects noticed earlier when looking at the 2S hyperfine splitting seem to be insignificant in the $2^3P$ hyperfine splittings. Therefore, some final conclusions can now be drawn about the actual $2^3P$ hyperfine splittings. Table 1 shows an uncertainty budget for these tests, while Table 2 lists the final results for these hyperfine splittings.

| *Source* | *Uncertainty (kHz)* |
|---:|---:|
| $1^{st}$ Order Doppler | 1 |
| Magnetic Field | 1 |
| Line Shape (step size) | 1.5 |
| Stopwire Alignment | 2 |
| Other | 1 |
| **Total (rms sum)** | 3 |

Table 1. Uncertainty budget.

| Hyperfine Interval | Measured Splitting (MHz) |
|---|---|
| **J = 0, F = 1/2 to J = 1, F = 1/2** | 28092.858(3) |
| **J = 1, F = 1/2 to J = 1, F = 3/2** | 4512.213(3) |
| **J = 1, F = 3/2 to J = 2, F = 5/2** | 1780.879(3) |
| **J = 2, F = 3/2 to J = 1, F = 1/2** | 668.007(3) |

Table 2. Final experimental results.

Comparison to Current Theory

The results for this experiment have been compared to the best theory to date for $^3$He. Although this theory is perfectly consistent with previous measurements, the new level of precision is an order of magnitude better, and this creates rather large discrepancies between theory and experiment. Attempts were made to fit several of the constants individually to force better agreement, with little success. Adjusting just one of the constants did very little to bring all of the intervals into agreement. The result of the comparison between the values from this experiment and theoretical values using the interaction constants calculated by Drake [6] are given in Table 3.

| Hyperfine Interval | Discrepancy (kHz) |
|---|---|
| **J = 0, F = 1/2 to J = 1, F = 1/2** | 17(3) |
| **J = 1, F = 1/2 to J = 1, F = 3/2** | -20(3) |
| **J = 1, F = 3/2 to J = 2, F = 5/2** | 1(3) |
| **J = 2, F = 3/2 to J = 1, F = 1/2** | 26(3) |

Table 3. Comparison to theory-discrepancies.

# CHAPTER 5

## CONCLUSION AND REMARKS

Though there are many challenges to the measurement of the hyperfine splittings in $^3$He, the results of this experiment in my judgment are reliable to the 2 kHz uncertainty quoted in the previous chapter. This is an order of magnitude better than any published measurements [3] for these splittings in previous experiments. While the theory is not in good agreement with these results, it is also clear that this theory has not been tested to this level of precision before. In fact, second order energy corrections, that have yet to be carried out, which involve cross terms between the hyperfine and fine structure interactions, could account for much of this discrepancy. In a private communication, Gordon Drake has informed me that his group is performing these very calculations. I'm certain the work being done in his group will be very enlightening to the understanding of these hyperfine splittings.

As for this work, there are certainly plans to improve upon the measurements discussed here. More time must be invested into closer examination of some of the systematic effects involved with performing these measurements. Also, at the time of this writing, my group is in the process of building a new apparatus with many significant improvements over the current setup. These include a significantly more stable and controlled magnetic field and an ultrahigh vacuum environment for much lower base pressures to decrease the background noise. Also, with the great success of the recirculation system develop for these $^3$He measurements, the new apparatus will operate exclusively in this mode for both $^3$He and $^4$He measurements. However, for the preparation of the initial states, a new approach is being considered. Instead of using an extra Stern-Gerlach magnet to compensate for the deflection of the metastable zero-states in $^3$He, these magnets will be done away with all together. The idea is to use a laser to pump the ±1 states into one of the zero states. Clearly this has the advantage of not only eliminating the deflection of the atoms, but also it will increase the number of zero-states and thereby the signal size. It will be very interesting to see the results from this new setup. So, there is no doubt plenty more to be done with these $^3$He hyperfine measurements.